\documentclass{nature}

\usepackage{verbatim}
\usepackage{times}
\usepackage{graphicx} 
\usepackage{chngcntr}
\usepackage{color}
\usepackage{booktabs}
\usepackage{amsmath}
\usepackage{url}
\usepackage{xcolor} 
\usepackage{comment}
\usepackage{hyperref}
\usepackage{cite}    
\usepackage{nameref}  
\usepackage{xr-hyper} 
\usepackage{makecell}
\usepackage{hhline}
\usepackage{lineno}
\usepackage{caption}

\usepackage[most]{tcolorbox}
\newtcolorbox{highlighted}{colback=yellow,breakable,lefthand width=0em,sidebyside gap=0pt,left=0pt,right=0pt,top=0pt,bottom=0pt,boxsep=0pt, boxrule=0.0pt,colframe=white!60}

\usepackage{xparse}
\usepackage{amssymb}

\ExplSyntaxOn

\keys_define:nn { miguel/label }
 {
  label   .tl_set:N = \l_miguel_label_tl,
  unknown .code:n   = \clist_put_right:Nx \l_miguel_label_clist
                       { \l_keys_key_tl = \exp_not:n { #1 } }
 }
\clist_new:N \l_miguel_label_clist
\box_new:N \l_miguel_label_box
\box_new:N \l_miguel_label_image_box

\NewDocumentCommand{\xincludegraphics}{O{}m}
 {
  \tl_clear:N \l_miguel_label_tl
  \clist_clear:N \l_miguel_label_clist
  \keys_set:nn { miguel/label } { #1 }
  \tl_if_empty:NTF \l_miguel_label_tl
   {
    \miguel_includegraphics:Vn \l_miguel_label_clist { #2 }
   }
   {
    \hbox_set:Nn \l_miguel_label_image_box
     {
      \miguel_includegraphics:Vn \l_miguel_label_clist { #2 }
     }
    \hbox_set:Nn \l_miguel_label_box
     {
      \skip_horizontal:n { 3pt }
      \fcolorbox{white}{white}{\footnotesize \tl_use:N \l_miguel_label_tl}
     }
    \leavevmode
    \box_use:N \l_miguel_label_image_box
    \skip_horizontal:n { -\box_wd:N \l_miguel_label_image_box }
    \hbox_overlap_right:n
     {
      \box_move_up:nn
       {
        \box_ht:N \l_miguel_label_image_box - 
        \box_ht:N \l_miguel_label_box - 3pt
       }
       { \box_use_drop:N \l_miguel_label_box }
     }
    \skip_horizontal:n { \box_wd:N \l_miguel_label_image_box }
   }
 }

\cs_new_protected:Nn \miguel_includegraphics:nn
 {
  \includegraphics[#1]{#2}
 }
\cs_generate_variant:Nn \miguel_includegraphics:nn { V }

\ExplSyntaxOff

\usepackage{filecontents}

\hypersetup{  
    colorlinks=true,
    linkcolor=blue,
    filecolor=magenta,      
    urlcolor=cyan,
    pdftitle={Sharelatex Example},
    bookmarks=true,
    pdfpagemode=FullScreen,
    citecolor=blue
    }

\title{A physically cryptographic warhead verification system \newline  using neutron induced nuclear resonances}

\author
{Ezra M. Engel, Areg Danagoulian$^{\ast}$\\
\\
\normalsize{Department of Nuclear Science and Engineering, Massachusetts Institute of Technology,}\\
\normalsize{77 Massachusetts Avenue, Cambridge, MA 02139, USA}\\
\\
\normalsize{$^\ast$Correspondence and requests for materials should be addressed to
A.D. (email: aregjan@mit.edu)}
}

\date{}

\begin{document}


\baselineskip18pt  

\maketitle

\section*{Abstract}

\begin{abstract}
Arms control treaties are necessary to reduce the large stockpiles of the nuclear weapons that constitute one of the biggest dangers to the world. However, an impactful treaty hinges on effective inspection exercises to verify the participants' compliance to the treaty terms. Such procedures would require verification of the authenticity of a warhead undergoing dismantlement. Previously proposed solutions lacked the combination of isotopic sensitivity and information security. Here we present the experimental feasibility proof of a technique that uses neutron induced nuclear resonances and is sensitive to the combination of isotopics and geometry. The information is physically encrypted to prevent the leakage of sensitive information. Our approach can significantly increase the trustworthiness of future arms control treaties while expanding their scope to include the verified dismantlement of nuclear warheads themselves. 

\end{abstract}

\section*{Introduction}


As of 2019 there are estimated 13000 nuclear weapons that make up the nuclear arsenals of the United States and Russia\cite{kristensen_US2018,kristensen_Russia2019}.  Such large arsenals may be one of the greatest threat to our civilization.  While high, these numbers are a significant reduction from the Cold War era, as a result of series of arms control treaties. The past treaties between United States of America and Soviet Union / Russia, however, primarily focused on the verified dismantlement of the delivery systems, such as ballistic missiles and strategic bomber aircraft.  This circumstance in part was driven by the notion that the delivery systems can be a good proxy for the states' strike capability during a nuclear war.  It was also driven by the difficulty of verifiably dismantling the warheads themselves without  leaking sensitive information about the weapon designs in the process\cite{ref:Hecker}.  This approach has left behind large stockpiles of surplus nuclear weapons, exposing them to the risk of theft and unauthorized or accidental use, as well as transfer to third countries.  This state of the affairs elevates the risk of nuclear terrorism and nuclear proliferation. The need for more effective arms control treaties is recognized by the 2018 US Nuclear Posture Review, which noted "The United States will continue its efforts to seek arms control agreements that enhance security, and are verifiable and enforceable."\cite{posture} Furthermore, there are increased worries that unless new treaties are implemented the current arsenal sizes will stagnate at the current numbers\cite{kristensen_Russia2019,kristensen2019global}. New treaties involving the verified dismantlement of nuclear warheads will significantly improve global security. However new types of technologies are necessary to enable such treaties. These technologies will have to detect hoaxing attempts, clear honest warheads as such,  while simultaneously protecting sensitive information about the weapon designs.

A variety of paradigms of verification have been proposed by a number of scholars and scientists.  The most recent of these, the template verification paradigm, is based on the expectation that the inspection party will be able to acquire an authentic device (henceforth referred to as the genuine reference object), and then use relevant encrypted data (known as a template) from this device to compare with equivalent data acquired from candidate devices of the same design.  The key to this verification process is a proof system that can guarantee that no treaty accountable item (TAI) undergoing verified dismantlement is secretly modified.  A number of works have been published on the philosophy behind this template verification - in the US national laboratories, think tanks, and academia\cite{fuller,drell1993verification,ref:sandia_exercise} - and a variety of concepts have been proposed  \cite{yan2015nuclear,ref:alex,philippe2016physical,PNAS,marleau2016implementation,gilbert2017single,ref:sandia_exercise,tris,vavrek2018experimental}.  Common to all these studies is the concept of the genuine reference object, which is acquired based on a situational context and brought to the testing facility via a chain of custody.  For a more detailed delineation of the procedure describing the acquisition of the genuine reference object (sometimes referred to as golden copy in the literature) and its use in the verification exercises see Results section in our previous work, Hecla and Danagoulian\cite{hecla2018nuclear}.  That work as well as the one presented in this study leverage Neutron Resonance Transmission Analysis (NRTA).  This technique uses neutron induced nuclear resonances in the electron-Volt (eV) energy range, common in many high Z elements including molybdenum, tungsten, uranium, and plutonium, to achieve high resolution imaging of the individual isotopes of those elements\cite{ref:chichester2012assessing}.  The strength of NRTA stems from the uniqueness of the resonance energies and amplitudes to individual isotopes, thus making their observation a unique tag of a particular isotope's abundance in the sample.  NRTA has been previously used to image and study various nuclear fuel samples and archaeological objects - for an in depth discussions see Losko et al. and Bourke et al.\cite{ref:losko,bourke2016non}.

Some of the previously proposed concepts made use of physical cryptography.  This has the advantage of avoiding computational cryptography, which can be prone to hacks and backdoor exploits and as such shifts the burden of verification to the computer code and electronic components. Work by Philippe et al.\cite{philippe2016physical} had the advantage of achieving the highest level of information security via a zero-knowledge measurement, yet it is not fully sensitive to isotopic hoaxes due to the similarity of fast neutron processes between various nuclei.  Additional work is currently underway to mitigate this circumstance, e.g. by observing fast neutrons from neutron-induced fission\cite{philippe2016inmm}.  Others, such as those by Vavrek et al.\cite{vavrek2018experimental} leverage nuclear resonance phenomena to achieve  strong isotopic sensitivity to hoaxes, but do not have zero-knowledge information security and involve complex systems that may be unpractical in a verification setting.  Our previous study, Hecla and Danagoulian\cite{hecla2018nuclear}, was based on Monte Carlo (MC) simulations, where we combined the advantages of a neutron-based radiographic system with the isotopic-sensitivity of a nuclear resonance sensitive measurement to achieve an effectively zero knowledge proof system.

\section*{Results}

\subsection{Template verification}

This work builds upon the general concepts described previously by Hecla and Danagoulian\cite{hecla2018nuclear}, with important differences and modifications.  The study by Hecla and Danagoulian was based on Monte Carlo simulations alone and envisioned an imaging system where the physical encryption is achieved by a geometrically complex reciprocal mask. The concept proposed in this work meanwhile achieves an experimental feasibility proof of concept involving a simple encrypting filter,  and uses a single pixel neutron detector.  The detector measures the combined flux through the object and can determine the energy of individual neutrons.  While a single pixel measurement along one axis of measurement does not have geometric uniqueness, such a uniqueness can be achieved via multiple measurements with object rotations, not dissimilar to the principles of a tomographic system.  This requires a new procedure for the comparisons between template data from the genuine reference object and the equivalent data from the candidate items.  This study's advantage is in the simplicity of the measurement as it avoids actual imaging and thus enables higher degree of information security.  It is also a direct experimental proof of feasibility which is sensitive to the backgrounds, noise, and system instabilities not captured in a numerical simulation.

\begin{figure}
    \centering
    \includegraphics[width=1\textwidth]{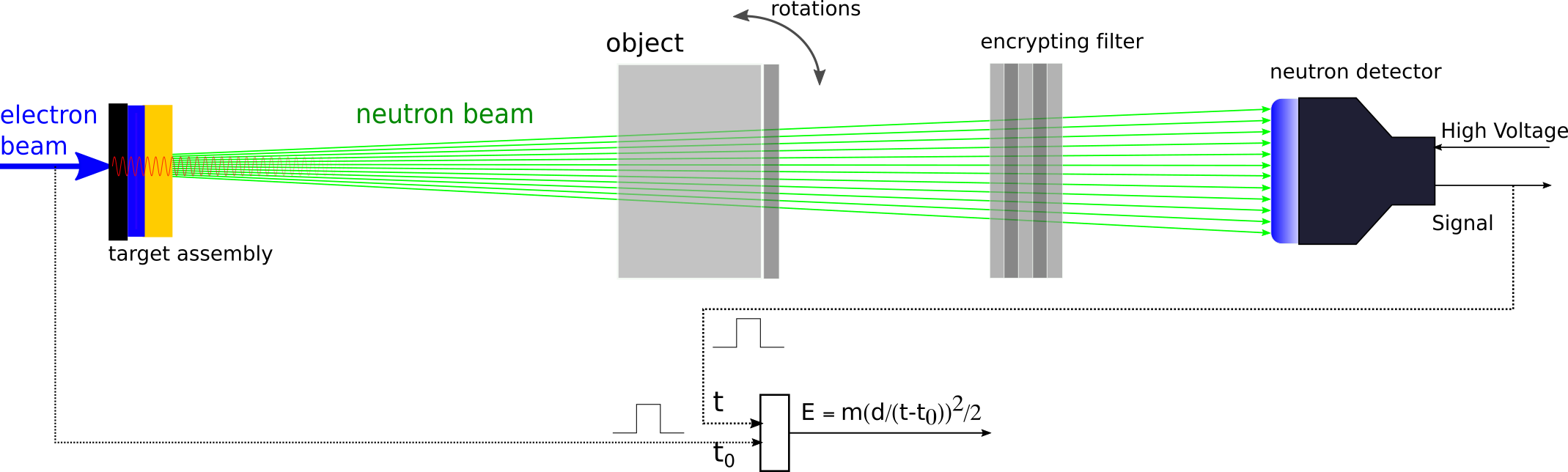}
    \caption{General schematic of the experiment (not to scale).  The combination of isotopic and geometric sensitivity is achieved by neutron spectroscopy via time-of-flight (TOF) technique and comparisons between the candidate and the genuine reference under random projections.  For simplicity, in this experiment the angles of the projections were chosen to be  0$^\circ$, 45$^\circ$, and 90$^\circ$.}
    \label{fig:drawing}
\end{figure}

The overall high level verification exercise is described in detail in Results section of Hecla and Danagoulian\cite{hecla2018nuclear}.  In brief, in an unannounced visit to an ICBM base the joint team of the inspectors and the hosts acquires a randomly selected nuclear weapon, whose authenticity is thus established by its situational context. The weapon is then transported under joint custody, i.e. in the presence of both parties and via application of tamper-proof seals and checks\cite{smartt2016overview}, to a facility where it is disassembled in a controlled environment such that no new objects are introduced or removed.  The fissile hollow sphere, from here on referred to as the pit, is removed and used as a genuine reference object. This study focuses on the stage involving the actual measurements test, where the genuine reference object is being compared to the TAI candidate objects.  The test needs to detect both isotopic hoaxes as well as geometric hoaxes, described later in this section.  The experimental procedure consists of the following steps:

\begin{enumerate}
    \item The genuine object, an encrypting filter, and the neutron detector are  aligned along the beam z-axis.  The encrypting filter is produced by the host, and its composition is unknown to the inspectors.  The system schematic can be seen in  Figure~\ref{fig:drawing}.  The filter can be made of any material, with the assumption but not the requirement that the hosts will add elements identical to those in the TAI, with the only requirement being that the filter allow a significant flux of neutrons to reach the detector for an agreed upon measurement time.  A spectroscopic measurement is taken, providing the template data (from here on referred to as template) that the next measurements will be compared to.
    \item Keeping the encrypting filter and the detector in their places the genuine object is replaced with the candidate object at the same orientation.  A measurement under experimental conditions identical to step (1) is performed.
    \item The data from the measurement of the candidate object is compared with the template in a statistical test.  If the test identifies a statistically significant difference then the candidate is declared a hoax.  If the new measurement passes the statistical test then the procedure moves on to the next stage, where measurements under random rotations are performed.
    \item  Measurements described in the steps (1) and (2) are now repeated with both genuine and candidate objects rotated at a random angle.  To keep the number of unknowns larger than the number of measurements a new encrypting filter is provided by the hosts.  Like before, the filter is kept in place during the two (genuine, candidate) rotated measurements.  Similar to step (3) the two measurements are compared in a statistical test.  If the measurements pass the statistical comparison then the candidate is declared isotopically and geometrically identical to the genuine, and is thus accepted as genuine as well.
\end{enumerate}

This work focuses on a sensitivity study, experimentally demonstrating the feasibility of the comparison.  The technique is shown capable of clearing objects identical to the genuine reference object and rejecting objects which are isotopically or geometrically different.  The verification exercise focuses on the weapons grade plutonium (WGPu) pit, primarily due to the weapon's hydrogenous components' near-opacity to epithermal neutrons.   HMX explosive, for example, has a mean free path of $\sim$1.8 cm for epithermal neutrons.  Public domain data on estimates of a Soviet thermonuclear design\cite{ref:fetter1990gamma} indicate that the explosive's thickness can be $\sim$6.5~cm, translating to an attenuation factor of about 1000.  While not impossible, a transmission measurement of weapons at a lesser degree of disassembly would require either more intense neutron sources, or longer measurement times.  This may become possible in the future, as the laser driven and other high intensity neutron sources currently under development become more applicable for field use\cite{roth2013bright,fernandez2017laser}.  It should be added that the plutonium pit may also contain other materials, such as coatings to prevent the air's corrosive effects on plutonium, as well as the beryllium neutron reflector shell which may be bonded to the plutonium shell.  The later has a mean free path of 1.3~cm for eV neutrons, and depending on its thickness its attenuation effects may translate to measurement times somewhat longer than those estimated in this work. This study thus focuses on an inspection scenario where the plutonium pit has been extracted in a controlled environment and undergoes verification, as described in  prior work\cite{hecla2018nuclear}.

\begin{figure}
    \centering
    \includegraphics[width=0.95\textwidth]{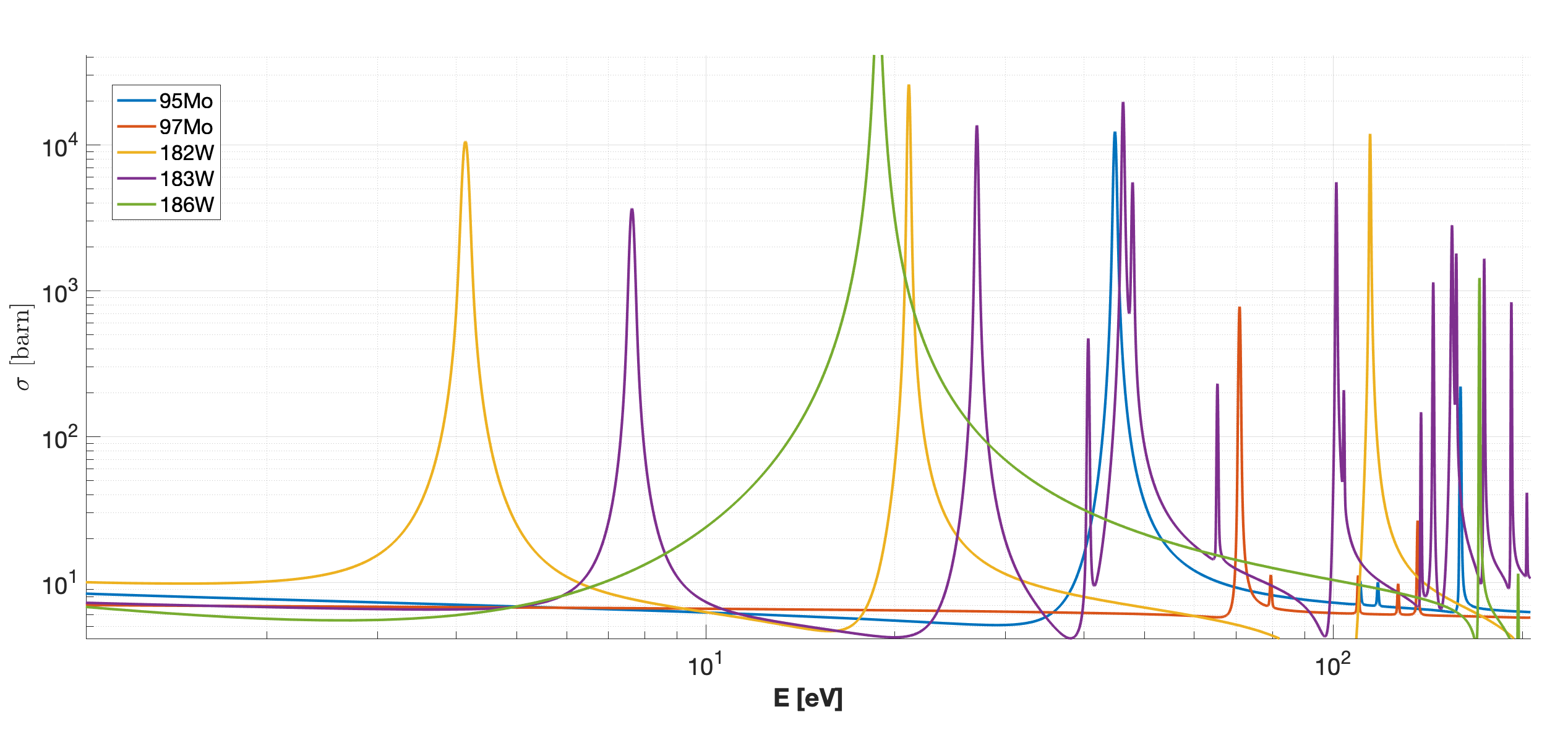}
    \caption{Cross sections.  Total interaction cross sections for neutron induced interactions for a subset of molybdenum and tungsten isotopes with significant resonances in the 1-200 eV range.}
    \label{fig:cross_sections}
\end{figure}

\begin{figure}
    \centering   
\includegraphics[width=1.\textwidth]{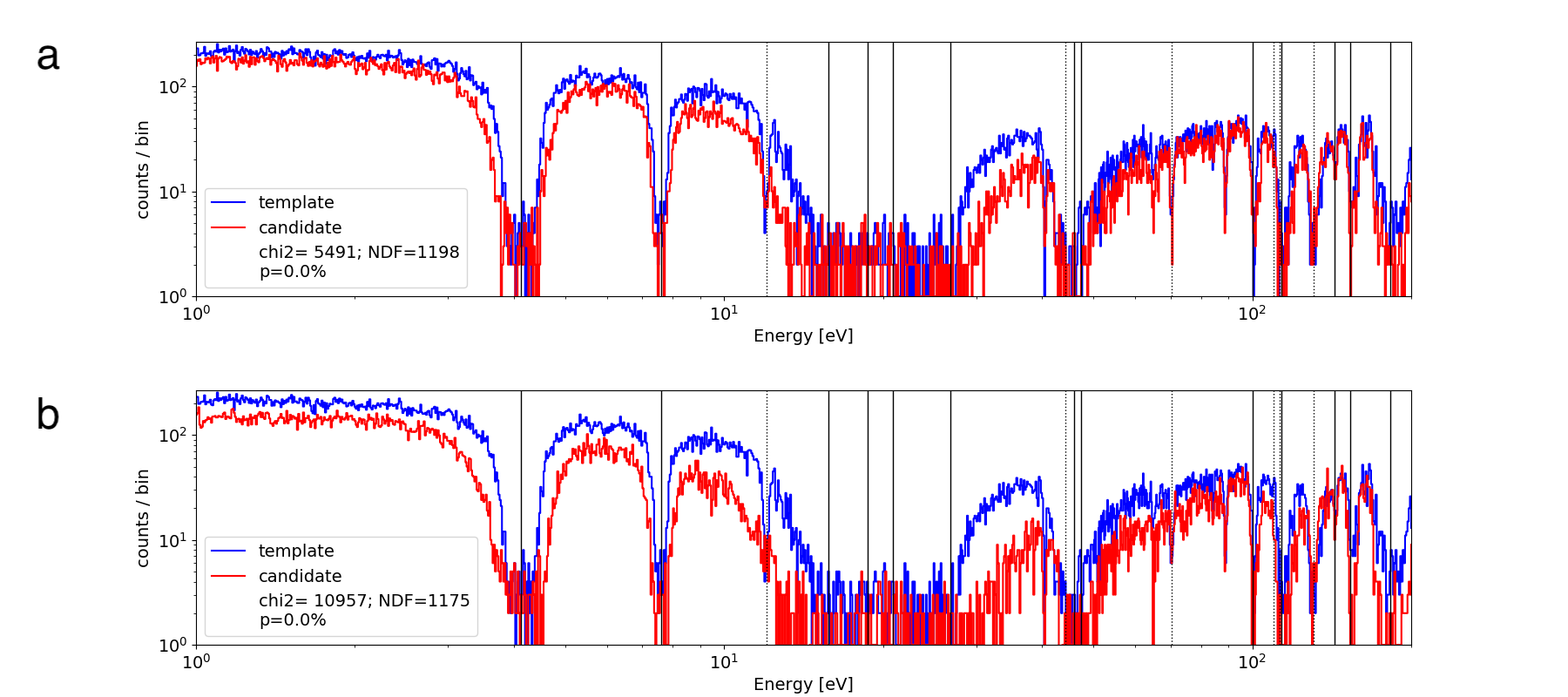}
    \caption{Histograms of neutron count energies for two isotopic hoaxes.  {\bf a}  The template data from an authentic reference of 90/10 Mo/W composition (blue) is compared to the data from a 50/50 Mo/W isotopic hoax (red).  {\bf b} The data from the authentic reference is compared to that from a 10/90 Mo/W isotopic hoax (red). The legend lists the $\chi^2$ value, the number of degrees of freedom, and the corresponding probability (p-value) for the $\chi^2$ test.  Both hoaxes are rejected.  The solid and dashed vertical lines denote the locations of some of the known tungsten and molybdenum resonances, respectively.   Data collection lasted approximately five minutes per object.}
    \label{fig:isotopic}
\end{figure}

\subsection{The experiment}

The overall diagram of the measurement setup can be seen in Figure~\ref{fig:drawing}, and is discussed in more detail in  ~\nameref{methods} section.  The  neutrons are produced in a pulsed mode from a pulsed electron linear accelerator, via a tantalum bremsstrahlung-photoneutron converter, which produces neutrons via the $(\gamma,n)$ reaction.  A 2.54~cm polyethylene moderator degrades the neutrons' energy from the $\sim$MeV scale to the $\sim$eV scale.  As the epithermal neutrons traverse the object and the encrypting filter, their spectrum is modulated in accordance to the cross sections and areal densities of the relevant isotopes, after which they are detected by a $^{6}$Li glass scintillator detector.  The detector produces a timing pulse, which is compared to the timing pulse of the linear accelerator. This comparison allows to determine $t_{\text{tof}}=t-t_0$ of the neutrons.  The energy of individual neutrons can then be reconstructed via $E=m(d/t_{\text{tof}})^2/2$, where $m$ and $d$ are the neutron mass and the flight path length, respectively.  The transmission spectra are normalized to the incident neutron count, determined by a fission chamber (not shown) upstream of the target object.

\begin{figure}
\centering
    \includegraphics[width=1\textwidth]{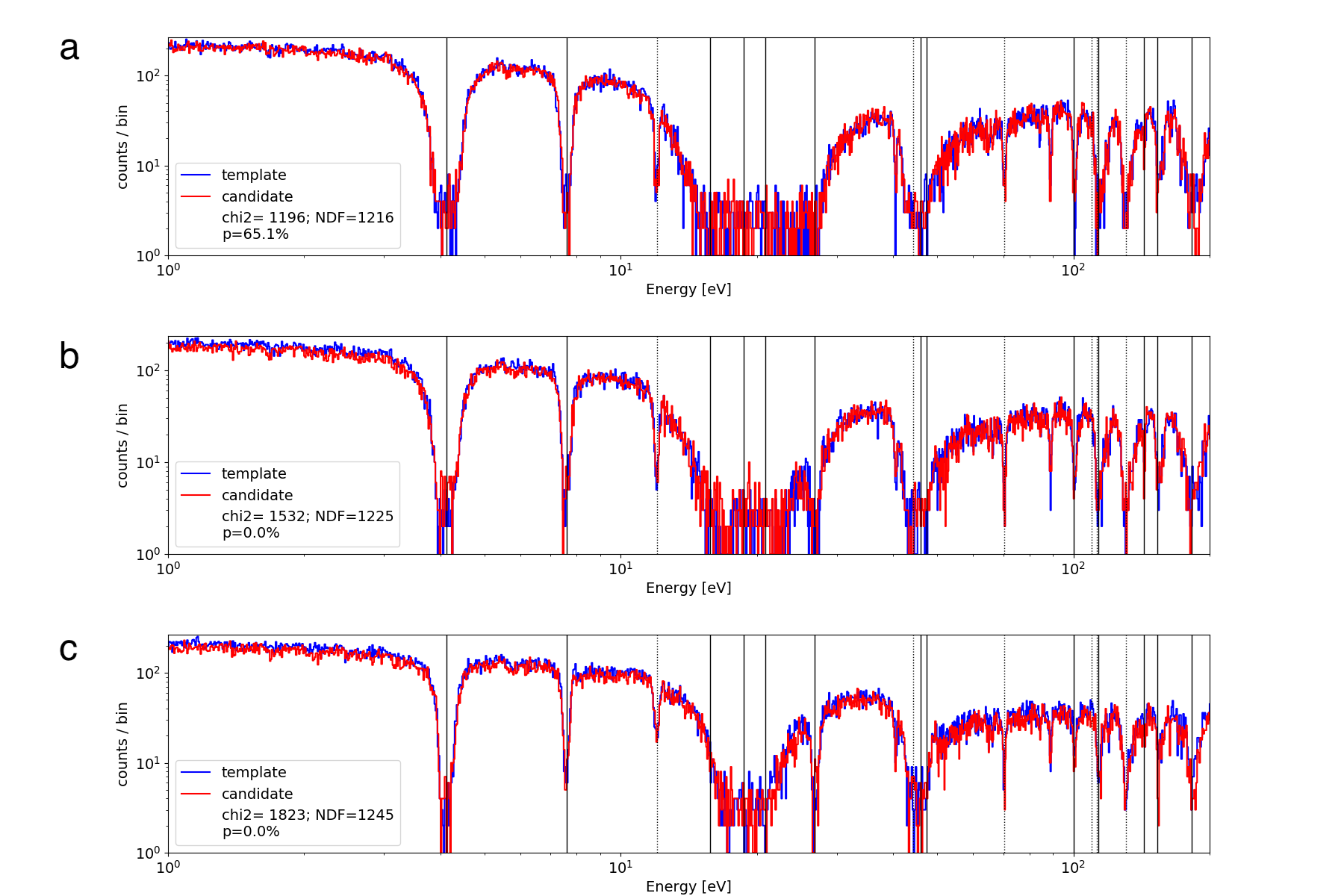}

    \caption{Histograms of neutron count energies for a geometric hoax.  The geometric hoax has been constructed as to have the same areal density along z-axis as the genuine reference.  The template data from an genuine reference (blue) is compared to the data from the geometric hoax (red) at $\theta=0^\circ$ ({\bf a}), $\theta=45^\circ$ ({\bf b}), and $\theta=90^\circ$ ({\bf c}).  As expected per design the $\theta=0^\circ$ yields a perfect agreement, as shown by the p-value.  The rotations readily expose the hoax.}
    \label{fig:geometric}
\end{figure}

\begin{figure}
\centering
\includegraphics[width=1\textwidth]{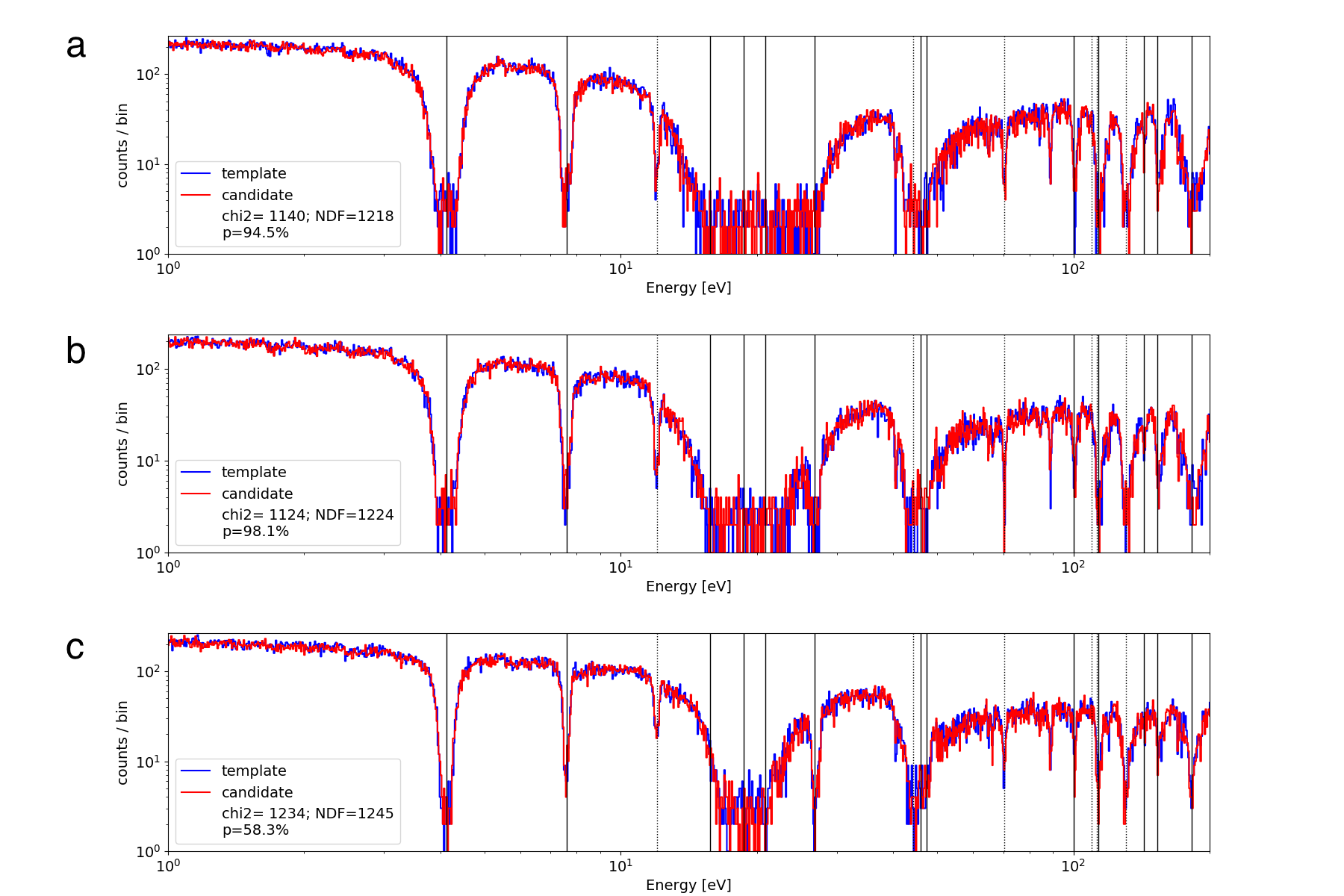}
    \caption{Spectral comparisons between a genuine reference  and an honest candidate.  The measurements were performed at rotation angles $\theta=0^\circ$ ({\bf a}), $\theta=45^\circ$ ({\bf b}), and $\theta=90^\circ$ ({\bf c}). All comparisons yield p-values of 50-95\%, indicating an agreement, and thus clearing the candidates as genuine objects.}
    \label{fig:honest_clear}
\end{figure}

\subsection{Targets}  As the research was performed in academic setting, plutonium and uranium targets were not available.  Thus, targets were built of more commonly available materials which have significant resonances in the relevant region of $1 \text{eV} \leq E \leq 200 \text{eV}$ and can thus act as proxies for most uranium or plutonium isotopes.  Natural molybdenum and tungsten were chosen as element-to-isotope proxies for $^{239}$Pu and $^{240}$Pu, respectively.  Instead of hollow spheres, which are typical to nuclear weapon pits, simpler cylindrical geometries were chosen, such that the genuine proxy object is a cylinder of 50.8 mm diameter and 30 mm length, with the first 27 mm consisting of molybdenum and 3 mm consisting of tungsten.  Such a combination was chosen to mimic the $\sim$90\% enrichment of WGPu. Plots of total neutron interaction cross sections for a subset of molybdenum and tungsten isotopes can be seen in Figure~\ref{fig:cross_sections}.  Unlike molybdenum or tungsten the actinide targets would produce additional neutrons via the (n,fission) reaction.  However the resulting neutrons would be of the $\sim$MeV energies, making the $^6$Li detector insensitive to them.  The encrypting filters  were assembled from 3~mm and 6.35~mm thick tungsten and molybdenum plates, respectively, that had 76.2~mm $\times$ 76.2 mm outer dimensions.  The measurements at 0$^\circ$ rotation, described in Figures~\ref{fig:isotopic},~\ref{fig:geometric}(a), and~\ref{fig:honest_clear}(a), used a filter consisting of three tungsten and three molybdenum plates aligned with the beam.  The measurements at 45$^\circ$ used a filter consisting of two tungsten and one molybdenum plates, and at 90$^\circ$ the filter consisted of one tungsten and one molybdenum plates. See~\nameref{methods} for additional detail on experimental settings. 

\subsection{Epithermal Neutrons}

In this experiment the neutron energies at the source spanned from the $~\sim$MeV scale, known as fast neutrons, to $\sim$meV, referred to as thermal neutrons.  However the use of a thin cadmium filter near the source filtered out most neutrons below 0.5 eV.  The detector used to register the neutron hits was based on $^{6}$Li isotope, which interacts with neutrons via the $^{6}$Li$(n,t)\alpha$ reaction, with the triton and the alpha causing scintillation in the glass and thus triggering detection.   When combined with the TOF technique this allows to determine the neutron's kinetic energy with $\ll$eV precision, allowing to resolve narrow absorption lines such as the 0.13~eV wide line at 70.9~eV from $^{97}$Mo.  The spectra in Figures~\ref{fig:isotopic},~\ref{fig:geometric}, and~\ref{fig:honest_clear} were produced via this method.

The results of isotope-specific resonant absorption are most readily observable in Figure~\ref{fig:honest_clear}.  The solid vertical gray lines indicate the expected positions of the tungsten resonances, while the dotted lines indicate the positions of the molybdenum resonances.    Depending on an isotope, the resonances are due to the (n,$\gamma$) and (n,n$'$) reactions, which selectively remove neutrons from the beam at the resonant energies. The relative depth of the absorption lines primarily depend on the resonance width, with broader widths translating to higher integrated cross sections and thus more absorption.  The shape of the absorption line in part depends on the type of the interaction.  These combined absorption features yield a unique spectral fingerprint of a particular configuration of isotopic, geometric, and density distribution.   Future research however should focus on providing a systematic proof of this uniqueness, i.e. that no combination of other elements can reproduce uranium or plutonium signatures.   See  Supplementary Note 4 for a detailed discussion. 

\subsection{Isotopic Hoax resistance}

Isotopic hoax resistance refers to a system's ability to detect hoaxing attempts which involve modifications of the isotopic ratios in an otherwise geometrically identical object.  For a real warhead this may involve replacing the WGPu pit, which has enrichment levels of more than 90\%, with a pit made from the more available Reactor Grade Plutonium (RGPu), which has a lowered $^{239}$Pu enrichment of approximately 60\%, with the remaining $\sim$40\% in the form of $^{240}$Pu.

To test the sensitivity of the technique to isotopic variations the 90\%:10\% Mo:W genuine reference object's transmission spectrum was compared to the transmission spectra of 50\%:50\% and 10\%:90\% Mo:W objects of the same cylindrical shape and overall dimensions.  Figure~\ref{fig:isotopic} shows the plots and comparisons of the spectra.  The comparisons show that as the proportion of tungsten is increased the tungsten absorption lines show increased absorption, while the molybdenum lines exhibit reduced absorption, as expected.  A simple $\chi^2$ test can be applied, determining the values of $\chi^2$  and the corresponding probability $p$ that the difference is merely due to random fluctuations.  For these comparisons the $p$-value is consistent with zero to ten significant digits, and thus indicates that a systematic difference is present.  The $\chi^2$  test thus rejects the comparison, and indicates a hoaxing attempt.  All measurements lasted approximately 5 minutes.

\subsection{Geometric Hoax resistance}

A single measurement along the beam's z-axis is only sensitive to the projected isotopics along that axis.  The transmission $T$ for a an energy bin $E$ and a spherically symmetric geometry can be approximated as 
\begin{equation*}
 \ln T \simeq -\sum_i \frac{\sigma_i(E) N_{\text{Av}}\rho_i}{A_i} 2\pi \int_{1}^{\cos \theta'} \int_0^X f_i(r,\cos \theta)  drd\cos\theta,   
\end{equation*}
where $\sigma_i(E)$, $\rho_i$, $X$, $\theta'$ and $A_i$ are the cross section, density, total thickness, polar angle, and atomic number for isotope layer $i$ of the object, respectively. $f_i(r,\cos\theta)$ is the fractional abundance of isotope $i$ at a particular radial coordinate, and $N_{\text{Av}}$ is Avogadro's number.  The summation takes place over all the isotopes, and the integration is along the radius and the polar angle.  Thus, the single pixel camera measurement effectively collapses the three dimensions into a zero-dimensional energy-dependent-only measurement.  This limitation can be readily overcome by making multiple measurements at various angles, similar to tomographic measurements where a single pixel can ultimately produce a three dimensional sensitivity.  Unlike a tomography where the goal is to produce a three dimensional image, here the goal is to simply verify that two objects are identical to within some level of geometric complexity.  Prior work has shown that two to three measurements at randomly selected angles are sufficient to achieve this goal\cite{PNAS}.  The random selection of the angles at the time of the verification exercise renders the hosts unable to optimize a possible geometric hoax to the measurement.  The problem is thus reduced to proving that the system is sensitive to geometric differences under rotations.

To show this sensitivity we perform measurements at 45$^\circ$ and 90$^\circ$ relative to the beam's z-axis.  For this purpose the geometric hoax object was chosen to be a rectangular parallelepiped, i.e. a box, of the same composition along the z-axis as the genuine reference object.  It consists of a 50.8 x 50.8 x 27 mm$^3$ box of molybdenum followed by a 50.8 x 50.8 x 3 $mm^3$ plate of tungsten.  Since the beam diameter is only 44.5 mm, the spectral measurement results from two objects should look identical at 0$^\circ$.  However at 45$^\circ$ and 90$^\circ$ the geometric differences should amount to different values of $\ln T$.

To test the above hypothesis and the sensitivity of the system under rotations the measurements were performed under identical beam conditions.  Furthermore, for every rotation the encrypting mask was modified but kept the same between genuine-candidate comparisons.  This is a necessary step, as to prevent any differential analysis of data from multiple angles which may otherwise reveal geometric information. In analogy to an under-defined system, every new measurement needs an additional unknown, in the form of a different encrypting filter. Figure~\ref{fig:geometric} shows the spectral comparisons.  It can be seen that at 0$^\circ$ the spectra are essentially identical, when compared with a $\chi^2$ test.  However, when rotated the small but significant differences in geometries give rise to detectable deviations.  While not as large as in the isotopic hoax, the differences are nevertheless statistically significant.  This result shows that for these experimental settings, object types, and measurement times the system can detect these geometric hoaxes with rotations by $\theta \geq 45^\circ$.

In this experiment all the rotations were centered on the geometric center of the object.  However, hollow spheres of the same thickness but different radii would produce identical signals under such rotations, due to their spherical symmetry.  This would result in an effective geometric hoax.  To break this symmetry the rotations should either involve a randomly chosen center, or be followed by random translations. Overall this study focuses on a limited set of hoaxing scenarios, and future studies need to focus on further extending these tests.  As with most cryptographic protocols, the concept presented in this work should undergo future tests and checks.  The use of the K-transforms in particular may prove promising in providing a proof of geometric uniqueness\cite{PNAS}.

\subsection{Completeness}

Finally, the test procedure needs to also show that it is able to accept honest candidates.  As with the geometric hoax sensitivity test, here the test needs to clear the honest candidates at 0$^\circ$, $45^\circ$ and $90^\circ$ rotations, as a way of confirming that the objects are identical both isotopically and geometrically.  We thus repeat the procedure described in the previous subsection.  This is necessary in order to rule out the possibility that the previously observed differences were merely the results of system instabilities, such as beam energy variations or detector gain shifts.
Figure~\ref{fig:honest_clear} shows comparisons where the genuine reference object measurements are compared to those from identical, honest objects.  Between these measurements the honest objects as well as the encrypting filters were replaced with other targets and filters and then were placed back for the comparison measurements.  A marker on the target holder allowed for the precise setting of the rotational angle.  In one case the accelerator had to be turned off and restarted due to beam instabilities.  As expected, and despite technical disruptions, in all cases the $\chi^2$ test indicates that the differences are merely statistical, as the $p$-value of the test indicates an agreement.  It is then concluded that the system is capable of clearing honest candidate objects.
A summary of the test results are listed in Table~\ref{tab:comparisons}.  Per procedure described at the beginning of the section, the test immediately rejects the attempts at isotopic hoaxing.  The geometric hoax shows an identical signal to the genuine reference object at measurements along the z-axis, however rotations by 45$^\circ$ or larger immediately results in a rejection.  Finally, the honest candidate measurements pass the test for all angles of rotation, and are thus accepted as honest objects.

It is necessary to acknowledge that a small rate of false positives may be caused by difference in ages between honest pits, caused by slight differences in $^{241}$Pu concentrations due its beta decay into $^{241}$Am, as well as machining tolerances in pit production.  As demonstrated in Supplementary Note 3 the former is a minor effect that should not cause false positives for pits with age difference of less than ten years. The later effect can be accurately estimated only given classified information on pit production, and should be addressed in the classified setting.  Our current modeling indicates that differences of less than 1~mm should not trigger false positives.

\begin{table}
    \centering
    \begin{tabular}{p{6cm} l  r r | r}
\makecell[l]{candidate object Mo/W\\ composition and shape} & \makecell[l]{rotation} & \makecell[l]{$\chi^2/$NDF} & \makecell[r]{p \\value} & decision \\
\hline
{\bf Isotopic Hoaxes}  &    &            &     &        \\
50/50, cylinder & 0$^\circ$  & 5491 / 1198 & 0.00 & reject \\
10/90, cylinder & 0$^\circ$  & 10957 / 1175 & 0.00 & reject \\ 
\hline 
{\bf Geometric Hoax} &    &           &      &        \\
               & 0$^\circ$  & 1196 / 1216 & 0.42 & accept  \\
90/10, cube     & 45$^\circ$ & 1532 / 1225 & 0.00 & reject \\
               & 90$^\circ$ & 1823 / 1245 & 0.00 & reject \\
\hline
{\bf Honest Candidate} &    &           &      &        \\
                & 0$^\circ$  & 1140 / 1218 & 0.94    & accept \\
90/10, cylinder & 45$^\circ$ & 1124 / 1224 & 0.98 & accept \\
                & 90$^\circ$ & 1234 / 1245 & 0.58 & accept
    \end{tabular}
    \caption{Statistical comparisons between the candidates and the genuine reference object. The composition here refers to the relative lengths along z-axis of molybdenum/tungsten components of the candidate objects.  The total length of the objects is kept identical to that of the genuine reference.  The genuine reference is a cylinder of the length of 3cm, and Mo/W composition of 90/10.  In order to pass inspection, the objects have to first pass the $\chi^2$ test along z axis ($\theta=0$), and then along two rotations at 45$^\circ$ and 90$^\circ$.  The procedure is shown to accept all honest candidates and rejects both isotopic and geometric hoaxes.}
    \label{tab:comparisons}
\end{table}

\subsection{Information Security}

One of the important goals of the verification system is the information security, that is the inspectors' inability to learn significant information about the TAI. In our prior work we showed via computational simulations that while the inspectors can learn information about the combined content of the TAI and the encrypting filter, they cannot learn anything specific to the TAI itself -- see Hecla and Danagoulian\cite{hecla2018nuclear}, subsection on Isotopic Information Security.  The primary focus of this study is the sensitivity analysis,  with the purpose of showing that the concept is capable of detecting hoaxing attempts.  Nevertheless, we use experimental data from the measurements in combination with a Geant4\cite{allison2016_geant4,mendoza2014new}  computational model of the experiment to simulate scenarios with a WGPu pit and an encrypting filter of WGPu weighting  1.8~kg.  In these data-driven simulations we show that  various opposite combinations of the TAI and filter geometries produce statistically indistinguishable signals. This makes it impossible for the inspectors to learn information of value, similar (but not identical) to the concept of zero-knowledge proof.  At the same time, additional simulations show that pairs of 5 minute long measurements at an experimental facility similar to the one used in this study can readily detect cheating scenarios where the WGPu pit has been replaced with RGPu, or where its size has been reduced by 2~mm or less, depending on experimental conditions.  See Supplementary Note 3 for a detailed discussion.  In Supplementary Note 1 it is demonstrated that using an encrypting filter of just 1.8~kg will result in an inferred range of pit's possible mass that spans from zero to a value that is significantly larger than the critical mass. At the same time the combined target thickness will be small enough to allow for significant counting statistics and thus reasonable measurement times.  The impact of temperature dependent broadening of the resonances is also estimated and described in Supplementary Note 2, showing that the effect can be made negligible.    

The presence of absorption lines associated with weapon materials whose presence may be a priori unknown to the inspecting side is an important consideration in the context of information security.  If those materials were to have resonances in the energy domain of the measurement, the inspector would learn about their presence.  Given that this study focuses on a scenario where a bare pit undergoes verification, the list of the elements that would be of significant quantities is limited to the actinides and gallium, which is used to stabilize the plutonium in its $\delta$-phase. Nevertheless, to address possible concerns, the treaty signatories can agree that the host will put samples of all elements that contain resonances in the agreed-upon range in their encrypting filters.  In this manner the inspectors will not be able to determine whether the absorption lines come from the weapon, or the filter.  The problem also has a technological solution.  A neutron-opaque chopper synchronized with the neutron pulse can be added to the beamline, with an opening chosen such that it only lets through neutrons of a specific energy range. For example, for this experiment the [1,50]~eV range translates to TOF of [150,1061]~$\mu$s.  With an accelerator pulse rate of 25 Hz, a chopper rotating at 12.5 Hz would need two 41$^\circ$-wide openings centered at 27.25$^\circ$ and 207.25$^\circ$ phases relative to the pulse to pass only the [1,50]~eV neutrons. This  opening angle and the rotation phase can be tuned to match the desired energy range,  including that of a single resonance.  Any hoaxing attempts involving shifting of phase or frequency would be readily exposed by the TOF data.

Finally, the chain of custody for the selection of the genuine object and its tamper-proof transportation to the dismantlement and verification facility is an important topic.  Past literature has discussed various approaches for achieving this goal\cite{fuller,bunch2014supporting,HassanElBahtimy}.  Exercises by the US and European national laboratories and organizations have taken place to test and characterize series of techniques necessary for this goal.  The Letterpress Exercise in particular has focused on the high level procedure, as well as the various simple but robust tamper-proof methods.  Such techniques include electrical, mechanical, as well as reflective particle tag seals.  Many of these low-tech methods use a physical randomness to ensure that a broken seal cannot be faked or reproduced.  For a detailed discussion see Smartt and Marleau\cite{smartt2016overview} and Bunch et al.\cite{bunch2014supporting}.  In addition to the particularities of the chain of custody, the arms control treaties stipulating verified dismantlement will have to negotiate the various conditions under which a particular verification technique is applied.  For the method described in this study in particular, the two sides should negotiate the following parameters:  the measurement time, beam current, and the resulting statistics, such that it is sufficient to detect the likely hoaxes without raising any worries about excessive information;  the minimum count rate to be observed by the detector, derived from broad considerations of the weapon material quantities, as a way of ensuring that the hosts do not add neutron-opaque materials to the encrypting filter and thus make the measurements impossible; the manner in which the detector and source instrumentation will be validated before and after the measurements, as away of ensuring that no hoaxing and no information leakage has occurred.  Generally, the use of physical cryptography makes the acquired data fundamentally encrypted, thus significantly reducing information security risks associated with the instrumentation that acquires the data.  However validation of the instrumentation can further strengthen confidence.  To this end, the inspection side could be asked to provide $n$ copies of the  instrumentation package, with the option for the hosts to perform invasive validation of the $n-2$ copies, leaving the last two copies to be used interchangeably in the verification exercises.  In this case the probability of the inspecting party to introduce two malicious packages undetected is $p= 1/ {n\choose 2} = 2/[n(n-1)]$. It should be clarified that we make no claims that the concept proposed in this study is a zero-knowledge proof.  Physical cryptography implies a degree of information security less than zero knowledge, with a goal of protecting most useful information, but not necessarily all information.  The considerations listed in this section are only part of a larger set of possible circumstances to consider. As is the case with most cryptographic schemes, future research should focus on the pursuit of possible vulnerabilities and for solutions for those.

\section*{Discussion}

Past arms control treaties lacked a method for reliably verifying the dismantlement of the nuclear weapons, and instead focused on the delivery systems, such as ICBM and strategic bomber aircraft.  This work experimentally demonstrates the feasibility of using resonance phenomena via epithermal neutrons transmission to compare two objects and verify that they are isotopically and geometrically identical, as part of a template verification exercise.  Any differences in isotopics and geometry between the candidate and the genuine reference object can be detected via isotope-sensitive measurements by using multiple rotations.  Similar to the requirements of zero-knowledge proof, the inspectors cannot learn significant information about the object itself, and at most can infer the combined mass and isotopics of the object and the encrypting filter. These combined inferred quantities thus constitute upper bounds for the object, corresponding to the scenario of an empty encrypting filter.  It is shown that by selecting thicker encrypting filters of just 1-2 kg of WGPu these upper bounds can be made large enough to make it impossible to infer any useful, new information.  Unlike most information barrier concepts the sensitive information is encrypted in the physical domain, thus achieving a very high degree of information security. For an in-depth discussion see the sections on information security in Hecla and Danagoulian\cite{hecla2018nuclear}. Using the neutron beam at an experimental facility similar to the one described in this study the measurements can be performed in the combined time of less than an hour. 

Additional and continued research in security studies arena and by the weapons laboratories is needed to address the various high level political and technical aspects of a treaty that would employ the techniques described in this study. In particular the joint chain of custody is paramount to the validity of the genuine object which establishes the template.  While no treaties stipulating the verified dismantlement of the weapons have been enacted, exercises by US and European national laboratories and institutes have explored the various real world considerations, testing a variety of protocols and techniques to insure a secure chain of custody\cite{smartt2016overview}. This study presents only a basic concept, which will need to be embodied in particular designs adapted to the particular treaty conditions and the characteristics of the available facilities and instrumentation.  

Finally, research needs to be performed to study the feasibility of this technique in more compact, possibly relocatable systems.  The work presented in this study used a large facility due to its convenience and the high precision neutron beam that it produced.  However smaller, more compact neutron sources may also enable such measurements.  Such sources could be in the form of a pulsed deuterium-tritium (DT) generator, or small, commercial, linear accelerator driven sources which use bremsstrahlung beams to produce neutrons via ($\gamma$,n) reactions.  The later systems are advantageous due to the large neutron flux that they would generate, and would be similar to the facility used for this work, albeit at much smaller, laboratory-size scales.  
While this study focused on the verification of plutonium components of a disassembled weapon, the advent of higher intensity neutron sources, e.g. using laser-driven methods\cite{roth2013bright,fernandez2017laser}, may allow for the verification of fully assembled weapons.  The use of compact neutron source platforms may also allow for implementations of the technique described in this study using non-electronic detectors and preloads, similar to the concept by Philipe et al\cite{philippe2016physical}, with the possibility of extending this physically cryptographic method to a fully zero-knowledge proof system.

\section*{Methods}\label{methods}

\subsection{Experimental Facility}

The experiment was performed at the 15~m station of the Gaerttner LINAC Research Center at the Rensselaer Polytechnic Institute. The neutron beam was produced via a 60~MeV electron beam from the LINAC, which produced 1$\mu$s-long pulses with a repetition of 25 Hertz.  An electronic signal from the pulse sets the start time $t_0$ necessary for the TOF calculation.  The beam was incident upon a tantalum photoneutron target.  Tantalum, due to its high atomic number, is a powerful source of bremsstrahlung photons when impinged by electrons.  At 60~MeV most of the energy loss by the electrons is radiative, i.e. via bremsstrahlung electromagnetic radiation.  At the same time the nucleus of $^{181}$Ta has a photoneutron production threshold of 7.6~MeV, via the $^{181}$Ta($\gamma$,n)$^{180}$Ta reaction.  The cross section for this reaction peaks to approximately 0.37~barns at 14.7~MeV.  Since the energy distribution of the bremsstrahlung photons is continuous, most of the photons overlap with the energy domain at which the photodisintegration of $^{181}$Ta is at a maximum.  The neutrons have an energy that is $E_\text{n} = E_\gamma-7.6$~MeV, where $E_\gamma$ is the energy of the bremsstrahlung photon.  Thus most of the neutrons are produced in the $\mathcal{O}$(MeV) range.  To produce an epithermal beam a 2.54~cm polyethylene target was placed near the tantalum target, allowing a small fraction of the fast neutrons to undergo elastic scattering on the hydrogen in polyethylene and thus lose energy in a process referred to as moderation.  The resulting neutron energy distribution spans thirteen orders of magnitude from fast, MeV scale to the thermal, meV scale.  For a detailed description of the facility and the target design see Section I in Danon et al.\cite{danon1995design}.

The resulting polyenergetic neutron beam was then transported via a beamline to the 15~m station, named so because of the target to detector distance of approximately 15~m.  A $^6$Li glass detector detected the neutrons, and registered their arrival times relative to the $t_0$ of the electron pulse.  The neutron detector consisted of  3~mm-thick lithium glass scintillator with ~6.6\% lithium content, enriched to 95\% in $^6$Li.  The scintillator is also sensitive to photons, however the photon arrival times would be $\mathcal{O}$(50ns), and thus could be rejected during the TOF analysis.  Additional photons are present during the neutron arrival times, primarily due to (n,$\gamma$) capture of the thermalized neutrons near the detector volume.  Those events however can be suppressed due to the relatively weak light output by photon interactions in the scintillator.  The only remaining background in the TOF analysis are the epithermal and thermal neutrons that scattered around the beamline.  This effect can be seen for example as the counts at the bottom of the 15-30~eV absorption lines in Figure~\ref{fig:isotopic}.  This background however is constant, doesn't change between multiple measurements, and thus doesn't affect the comparisons performed in this study.  For a detailed description of the detector, the data acquisition electronics, and the general experimental setup see Section II in Danon et al.\cite{danon1998neutron}.

\subsection{Energy reconstruction via TOF and statistical analysis}
  Due to a thermalization times of $\lesssim$ 1 $\mu$s and flight times of $\sim$ms, the pulse $t_0$ amounted to an good initial measure of the epithermal neutron emission time.  The detection of the neutron in the scintillator produces a second pulse with a time $t$, and the energy of the neutron can be then inferred from $E=m(d/t_{\text{tof}})^2/2$, where $t_{\text{tof}} = t-t_0$.  During the analysis of the data however it became apparent that the slight dependency of the moderation time on resulting neutron's energy significantly distorted the energy reconstruction.  This effect was particularly significant for the higher energy neutrons in the spectrum.  To quantify this effect simulations of this process were performed for the known geometries of the moderator, using the Geant4 object oriented simulation toolkit\cite{agostinelli_geant4}.  The energy-dependent corrections for thermalization time, which varied in the 4-8~$\mu$s range, were applied to the data, resulting in a precise reconstruction of the known molybdenum and tungsten resonances up to 200~eV. 
  
  The energy dependence of the detected neutrons in the open beam varied significantly with energy. This was caused in part by the intrinsic neutron flux, which changes significantly due to the incident neutron spectrum, caused by the underlying moderation process in the polyethylene moderator. It is also caused by the decreasing sensitivity of the $^6$Li glass detector with increased energy, due to the power-law dependence of the $^6$Li(n,t)$^4$He reaction, which has the form of $\sigma \propto E^{-1/2}$.  The data acquisition of the facility bins the counts in 0.1 $\mu$s wide time bins in the 1-200 eV range. When converting to the energy via TOF, the TOF distribution, which can be described as the time differential of the cumulative counting probability $dP/dt$, is weighted by the Jacobian, $dE/dt = -2E/t_{\text{tof}}$, such that $dP/dE = \frac{dP}{dt} / |\frac{dE}{dt}|$.  All statistical errors are propagated accordingly to the final energy spectrum. Finally, to even-out the statistics throughout the broad 1-200~eV range, the uniform binning is changed to a non-uniform binning, where increasingly larger bins are assigned with increasing neutron energy.
  
  The resulting spectral data from a pair of measurements were compared in a $\chi^2$ statistical test.  Assuming normally distributed errors, it is defined as $\chi^2 = \sum_1^N (c_{0,i}-c_{1,i})^2/(c_{0,i}+c_{1,i})$, where $c_0$ and $c_1$ are the count arrays for the spectra undergoing comparison, and $N$ is the total number of bins.  Given a value of $\chi^2$ and $N$, the probability, also known as p-value, can be determined from the $\chi^2$ distribution.  p-value describes the probability that two spectra's differences are due to statistical fluctuations alone.  A low p-value, e.g. $\ll$1\%, is indicative of statistical effects due to cheating by the hosts.

\subsection{Data availability}
All the data supporting the findings of this study are available from the corresponding author upon reasonable request.



\section*{References}

\bibliography{\jobname}

\bibliographystyle{naturemag} 

\clearpage

\section*{Acknowledgments}

This work is supported in part by by the Consortium for Verification Technology under Department of Energy National Nuclear Security Administration Award DE-NA0002534.  The authors are grateful for the support and encouragement from their colleagues within the Consortium of Verification Technologies.  The authors thank the staff at The Gaerttner LINAC Research Center and Yaron Danon in particular for assistance with the experiment, as well as valuable suggestions and advice. The authors are grateful to Igor Jovanovic for valuable comments on the manuscript.  The authors also thank Ethan A. Klein for cataloging the nuclear data for (n,$\gamma$) and (n,n$^\prime$) reactions on molybdenum and tungsten.  We express our gratitude to Jacopo Bongiorno for help with equilibrium heat transfer calculations, and to the reviewers for their insightful comments.

\section*{Author Contributions}

A.D. conceived the project.  A.D. and E.M.E. performed the experiments and analyzed the data.  A.D. wrote the manuscript.

\subsection{Competing Interests:}
The authors declare no competing interests.



\end{document}



\appendix
\baselineskip18pt
\section*{Supplementary Note 1} \label{sec:IS} 

\subsection{Uncertainty range in the inferred pit mass.}

Here we consider the scenario where an inspector uses the transmission data from a neutron beam to infer the range of possible pit masses.  To frame the problem, let's assume a hollow pit of Weapons Grade Plutonium (WGPu) in a $\delta$-phase of density $\rho \approx 15$g/cm$^3$.  We use public domain information about a Soviet thermonuclear warhead, as reported by Fetter et al.\cite{ref:fetter1990gamma}, which describe a hollow sphere of estimated inner radius of 6.27 cm and an outer radius of 6.7~cm.  Using a neutron beam of a radius of 2.5 cm (similar to the settings used in the experiments in this study), the hosts use a WGPu cylindrical encrypting filter of 2.5 cm radius and the length of 6 cm along the beam axis. 

Both the pit and the encrypting filter have to be presented for measurements in boxes made of materials that are optically opaque but transparent to neutrons, such as aluminum.  The box size could be about 20~cm.  Both the pit and the reciprocal are aligned with external fiducial markers on the box exterior, thus allowing for those same markers to be used in beam-target alignments -- for more detail see Hecla et al.\cite{hecla2018nuclear}, Supplementary Note 3.

We now assume that the inspector has analyzed the data, has determined the attenuation $A$ and has used $\ln A = -\mu d$, where $\mu$ is the linear attenuation coefficient, to infer the total thickness $d$ of the plutonium.  The inspector has then inferred,  correctly, that exactly 6.86~cm of WGPu are present along the z-axis of the beam.  What are then the possible values of the pit mass that the inspector can infer from these numbers?  There are two limiting scenarios that define the lower and upper bounds on the estimates:

\begin{itemize}
    \item[] {\bf Scenario 0}  At one extreme, the null scenario assumes that all the mass is found in the encrypting filter.  Thus for this limit the lower bound on the pit mass is $M_{\text{null}}=0$.
    \item[] {\bf Scenario 1} Here the inspector assumes the opposite extreme - that all the WGPu mass is in the pit and that the encrypting filter is empty.  In this scenario  
        the pit is a hollow sphere, with an outer diameter matching that of the box, i.e. 20cm.  In this case the mass of the pit is inferred to be $M_{\text{1}} = 15\cdot (4\pi/3) [10^3-(10-6.83/2)^3] $~grams, i.e. $M_{\text{1}}\approx$45~kg. 

\end{itemize}

Thus the inspector can only infer that the range of the pit mass is in the range of [0,45]~kg.  This constitutes useless information, as the critical mass of a bare plutonium sphere is $\sim$10~kg\cite{ref:explosive_properties}.  In other words the inspector has inferred numbers that range from the trivial (zero) to absurdly trivial (super-critical). 
In the discussion above we made an assumption that the transmitted signal is only dependent on the total areal density of plutonium traversed by the beam.  In \nameref{sec:SN3} we use experimental data and Monte Carlo simulations to prove this, showing that various geometric and isotopic pit-filter combinations produce the same signal, as long as the average enrichment and the total combined areal density is kept the same.  

In this process the host used an encrypting filter of cylindrical shape of a mass of approximately 1.8~kg.  An object of this mass is fully sub-critical, and should not be of any safety concern.  In a realistic verification campaign the host will need to produce filters of many thicknesses, furthermore using different thicknesses for different rotations.  However this does not mean that the host needs to produce a number of monolithic filters equal to the number of all weapons undergoing dismantlement times the number of random rotations.  The filter may consist of thin disks of plutonium grouped together, allowing the host to modify the thickness by changing the number of the disks.  For example, the host can produce 59 disks of 1~mm thickness, 99 disks of 10~$\mu$m thickness, and 10 foils of 1~$\mu$m thickness, allowing a range of thicknesses in the range of [0,6] cm, with a discretization of just 1~$\mu$m. As the protocol cannot resolve thicknesses of $\ll$0.1 mm, this discretization amounts to a continuum, allowing for continuously random filter thickness selections.  This single set of disks can be reused between measurements, albeit with different number of disks making up the filter, and will have a total weight of just 1.8~kg of WGPu.

It is necessary to point out that if a 6~cm filter is used, then the inspectors may be able to infer that the pit is not a solid sphere. The diameter of a solid sphere of 3.4~kg, the mass of the example used in this study, is closer to 7.6~cm.  To avoid such outcome the host simply needs to increase the thickness of the filter from 6~cm to 6.8~cm.  This would require approximately 10\% longer measurement time to achieve the equivalent statistics.    

\begin{figure}
{\bf a} \\
    \includegraphics[width=1\textwidth]{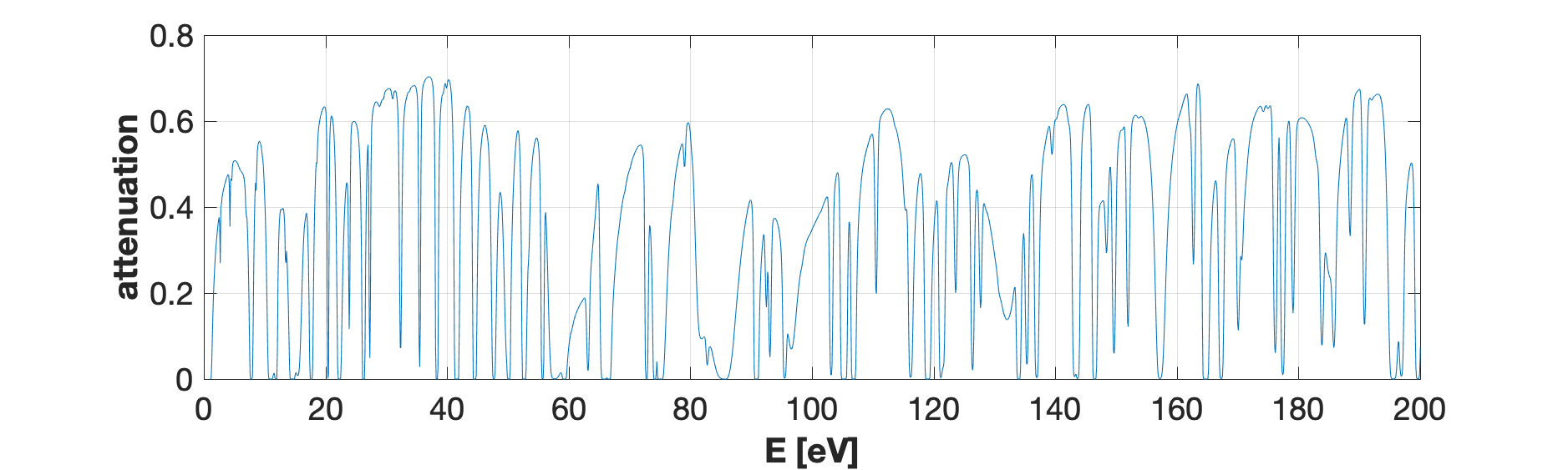} \\        
{\bf b} \\
\includegraphics[width=1\textwidth]{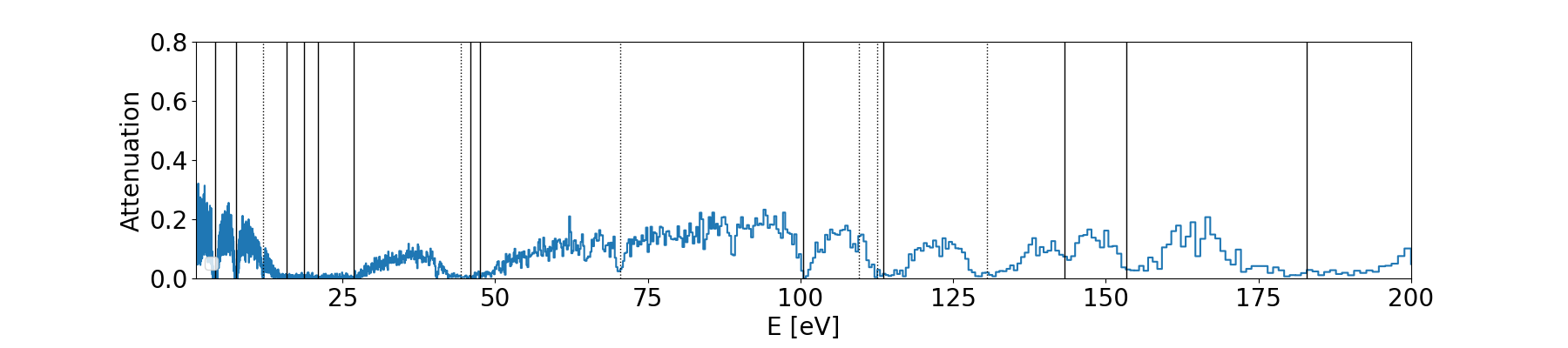}

    \caption{Dependence of transmission on the neutron energy.  Calculated for ({\bf a})  6.86~cm of WGPu via analytical calculations based on JEFF-3.2 cross section data, and ({\bf b}) for one of the genuine object samples in this study via experimental data.  By comparison to this study the realistic WGPu measurement scenarios will involve larger statistics and thus will involve shorter measurement times.}
    \label{fig:transmission}
\end{figure}

As a final consideration it is also important to determine the expected attenuation of the neutron flux by the combined WGPu thickness of 6.86~cm.  Using JEFF-3.2 cross section data it is possible to calculate the energy dependent transmission by this filter. Supplementary Figure~\ref{fig:transmission} plots this quantity, showing an average transmission of the order of $\sim$0.2.  For comparison, the targets used in the experiments described by this study involved transmissions significantly lower than this number.  It is then possible to conclude that a target of this combined thickness will still allow for collection of the necessary statistics in reasonable times, e.g. $\sim$5~minutes.  In \nameref{sec:SN3} we take this analysis further, using experimental data and Geant4 simulations for a WGPu pit geometry to show that 5~minute long measurements are sufficient to detect a hoax.

\section*{Supplementary Note 2} \label{sec:SN2}

\subsection{Effects of decay heat on Doppler broadening of the resonances}

Most of the isotopes in the plutonium pit of a warhead undergo alpha decay, which leads to constant generation of thermal energy.  This thermal energy results in a steady state heating of the pit.  Different levels of heating in the pit and the encrypting filter may give rise to a difference in resonant line widths due to different extents of Doppler broadening.  This could, theoretically, give rise to opportunities of inference of pit material quantities via careful fitting of the absorption lines by the appropriate functional forms.  It is possible to determine the extent of this heating, the steady state temperature of the pit, the extent of Doppler broadening, and thus describe the extent of this possible information vulnerability.

The WGPu contains about 94\% of $^{239}$Pu, as reported by Mark et al.\cite{ref:explosive_properties} in Table 1.  The nucleus undergoes alpha decay with the half life of 24110 years, emitting an alpha of 5.2~MeV.  One can then determine the total power of the heating, from
\begin{equation*}
    P = E\cdot 0.00012 \cdot \frac{ f M N_{\text{Av}}}{A} \ln(2)/\tau_{1/2}
\end{equation*}
where $E$ is the energy of the alpha, $f$ is the enrichment of a particular isotope, $M$ is the mass of the pit (3.4 kg for the example in \nameref{sec:IS}), $\tau$ is the half life of the isotope.  For $^{239}$Pu we find $P_{239} = 6$ Watt.  Adding decay from $^{238}$Pu and $^{240}$Pu the total power is $P_{\alpha} \approx 7$~Watt.

Approximating the pit as a homogeneous sphere of a constant volumetric heat production, we can now determine the steady state temperature at the surface of the pit, which is determined by the following equation\cite{bergman2011fundamentals}:
\begin{equation*}
    T_0 - T_{\infty} = qR/3h
\end{equation*}
where $T_{\infty}$ is the room temperature, $q$ is the volumetric heat production, $R$ is the outer radius of the pit, and $h$ is heat transfer coefficient, which is approximately 20 W/m$^2$/K for air flowing at 1 m/s, a speed that can be maintained with  a small fan installed in the opaque box containing the pit.  We then find that the surface of the pit will be only 6-7 K warmer than the surrounding air, translating to just 2\% difference in temperatures.  With plutonium's heat conductivity at $\sim$6 W/m/k, it can be shown that the pit will be at approximately uniform temperature.

Similarly, we can determine the average temperature of the WGPu encrypting filter as described in \nameref{sec:IS}.  For a cylinder of length $z$, radius $R$, and a total heat production rate $P$ the steady state temperature is
\begin{equation*}
    T_0 - T_{\infty} \approx \frac{P}{h(2\pi Rz+2\pi R^2)},
\end{equation*}
where $h=20$ W/m$^2$/K is the heat transfer coefficient of air flowing at 1 m/s\cite{bergman2011fundamentals}. Given the cylinder's size, it is possible to show that the Biot number is $Bi \ll 1$, allowing for the above approximation. Here we would like to point out that the encrypting filter does not have to be monolithic.  To address worries of criticality and to improve heat removal the 6 cm long cylinder can be split into four identical 1.5~cm long, 450~gram cylinders, aligned with each other and with the beam.  For one of these cylinders the temperature difference with the ambient temperature is then $T_0-T_{\infty} = 7$~K, which is less than 1~K higher than the temperature of the pit itself.

The Doppler broadening function for the resonance lines has a Gaussian form, with a standard deviation of $\Delta = 2\sqrt{\frac{E k_{\text{B}} T}{A}}$, see H\'ebert, page 35\cite{hebert2009applied}.   The intrinsic resonances are typically $\mathcal{O}(0.2eV)$ wide, while $\Delta\sim0.2$eV.  To understand its dependence on small variations in temperature, we can write $\Delta \propto \sqrt{T} \approx T_0 (1 + \frac{\delta T}{2T_0})$, where we have Taylor expanded the square root around $T_0=306.5$~K, the average temperature between the pit and the encrypting filter.  We then conclude that the $\delta T =\pm 0.5$~K difference in temperatures of the pit and the filter will translate to thermal broadening differences of $\sim \pm 0.08$~\%. It is then reasonable to assume that the difference in the total line widths between the pit and the encrypting filter is too small to be detectable in a measurement whose precision is fundamentally limited both by statistics and energy resolution.

However, if any doubts persist about differences larger than the one determined above, then the host can resolve this  by simply  mounting heating tapes to either one of the two objects, and perform temperature-controlled heating as a way of guaranteeing that the two objects are in fact at exactly the same temperature.

\section*{Supplementary Note 3} \label{sec:SN3}

\begin{figure}
    \centering
    \includegraphics[width=0.9\textwidth]{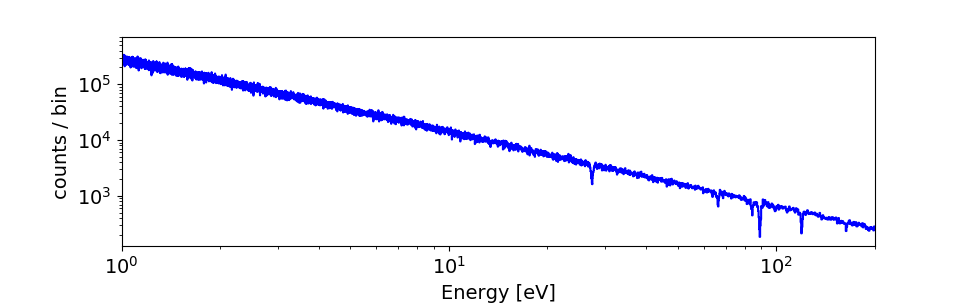}
    \caption{Spectrum of open beam neutron data from the experiment.  The observed absorption lines correspond to the resonances of the isotopes of a thin cadmium foil placed near the photoneutron target for filtering out the thermal neutrons.}
    \label{fig:open_beam}
\end{figure}

\begin{figure}
    \centering
    \includegraphics[width=0.9\textwidth]{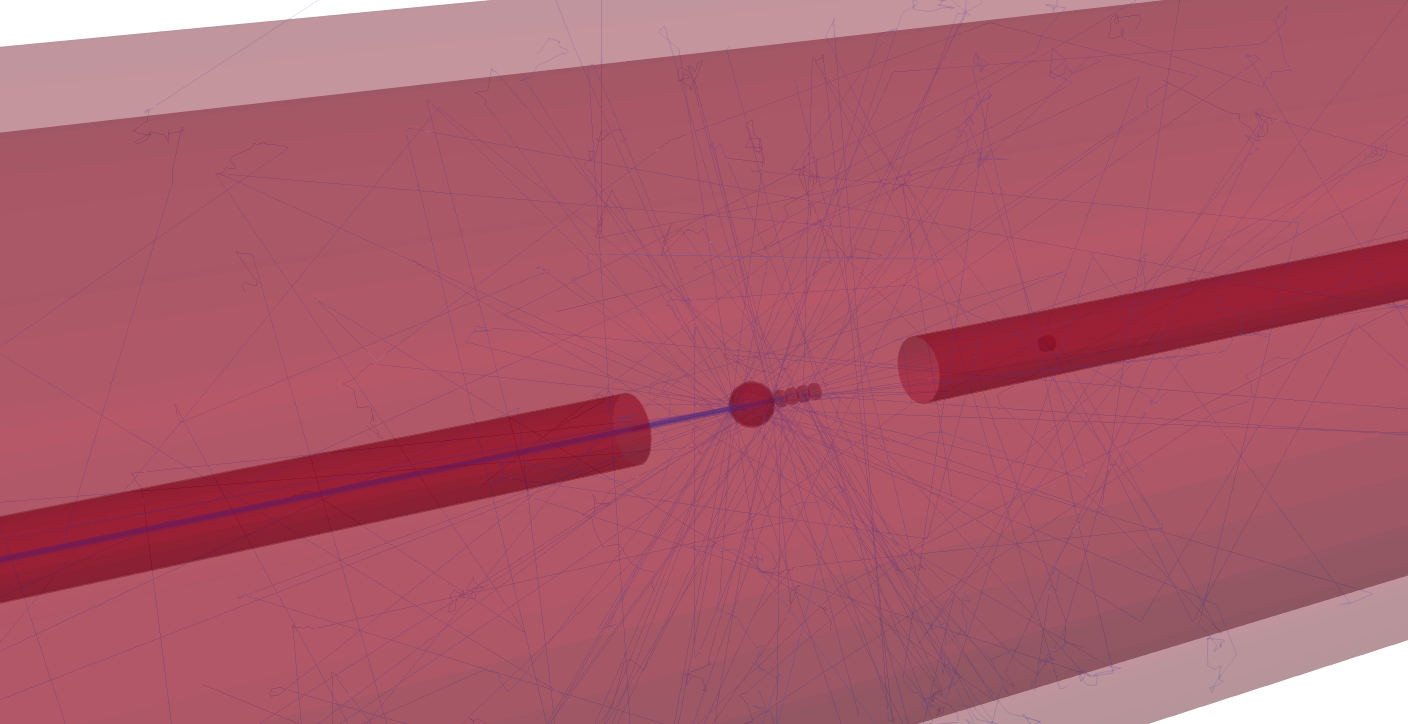}
    \caption{Rendering of the Geant4 model. It shows the beam line, the WGPu pit, and the encrypting filter.}
    \label{fig:rendering}
\end{figure}

The concept of Zero Knowledge proof requires that the verifier learns nothing new about the object undergoing verification.  In the context of an inspection exercise, it is important to show that the inspectors will learn nothing of value as a results of their observations.  In our prior work\cite{hecla2018nuclear}, Section Isotopic Information Security, we showed that the inspectors will be unable to make any isotopic inferences about the Treaty Accountable Item (TAI).  The experiments performed in this study focused on a sensitivity analysis, with the goal to show that hoaxes can be detected, and thus did not involve additional measurements testing the physical cryptography of the test.  However open beam data with no target or encrypting filters was collected. A Geant4\cite{agostinelli_geant4} computational model of the experimental facility uses this data as an input, allowing to simulate the transmitted neutron signal for the weapon pit and encrypting filter geometries described in \nameref{sec:IS}.  For simplicity, only the $^{239}$Pu and $^{240}$Pu isotopes were used.  The Geant4 used the JEFF-3.2 neutron libraries\cite{mendoza2014new}.  Supplementary Figure~\ref{fig:open_beam} plots a spectrum of the open beam neutron data.  The data shows that 821504 neutron counts were registered in the 5~minute long measurement. Supplementary Figure~\ref{fig:rendering} shows a visualization of the Geant4 model.  The $^6$Li glass detector used in this experiment was placed 14.573 meters from the photoneutron target, with the targets placed 110~cm upstream from the detector.  The flight path was located in a 2~m wide concrete tunnel. The 6~cm encrypting filter is split into four 1.5~cm section, for reasons described in \nameref{sec:SN2}.    Since the data already encodes the detector's intrinsic energy-dependent efficiency the detector in the simulation is modeled at 100\% efficiency.  The experimental data is histogramed in Supplementary Figure~\ref{fig:open_beam}.

\begin{figure}
{\bf a} \\
    \includegraphics[width=1\textwidth]{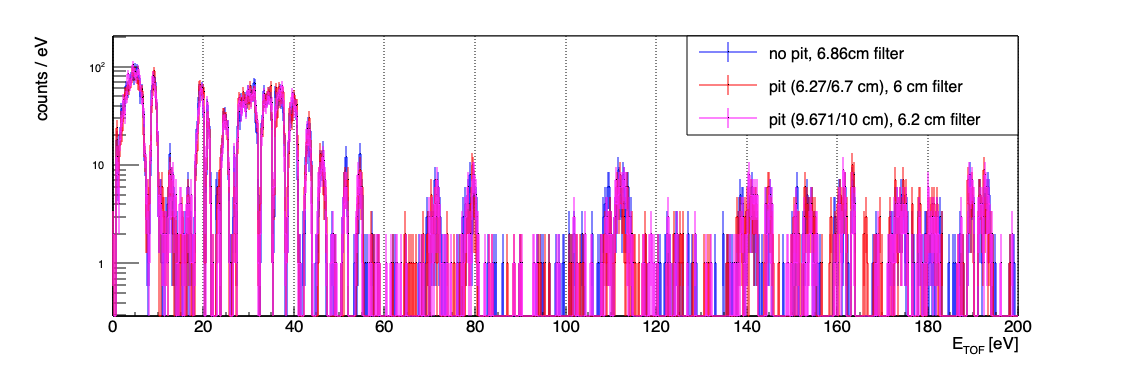}
{\bf b} \\
    \includegraphics[width=1\textwidth]{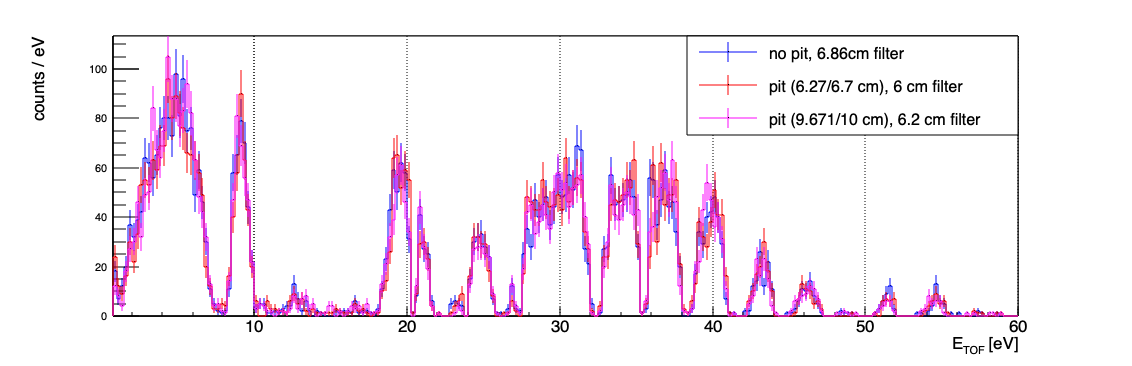}
    \caption{Results of the simulations of various pit masses.  The spectra for three scenarios are overlaid for 0-200 eV ({\bf a}) and 0-60 eV ({\bf a}) ranges, showing overall statistically identical distributions.  The $\chi^2$ test results can be found in Supplementary Table~\ref{tab:materials}.}
    \label{fig:materials}
\end{figure}

The main arguments presented in \nameref{sec:IS} are based on the assumption that the transmitted neutron signal is only sensitive to the total combined thickness or areal density of the pit and the encrypting filter.  This implies that the total transmitted signal cannot be used to infer specifics about the pit or the filter alone.  To prove this, we have ran a series of simulations where we vary the pit and filter dimensions and isotopics such that the combined areal density and the combined isotopic enrichment is kept constant.   In all simulations 821504 neutron counts were incident, sampled from the experimental data plotted in Supplementary Figure~\ref{fig:open_beam} and corresponding to a 5 minute measurement at the experimental facility used in this study. Supplementary Figure~\ref{fig:materials} shows the spectral results of the simulations in the 0-200 eV and 0-60 eV ranges.  Supplementary Table~\ref{tab:materials} lists the various scenarios and the results of the output comparisons in a $\chi^2$ test.  All of these have the commonality of combined areal density as traversed by the beam.  The $\chi^2$ tests show that the results for the various combinations produce statistically identical spectral signals, indicating that the signal cannot be used to infer information about the pit alone.  

\begin{table}
\label{tab:my-table}
\begin{tabular}{lllllllllll}
\hline
description &  radii, inner/  & $d_{\text{filter}}$ & total & pit  & \multicolumn{3}{l}{\makecell[c]{0-200 eV}} & \multicolumn{3}{c}{ \makecell[l]{0-60 eV}} \\
 &  outer [cm] & [cm] &  [cm] &  [kg]  & $\chi^2$ & NDF  & $p$ &  $\chi^2$ & NDF & $p$ \\ \hline
no pit &  0 / 0  & 6.86 & 6.86 & 0 &   -   & -  &  -    & - & - & -  \\
\makecell[l]{Black Sea\\model} & 6.27/6.7   & 6 & 6.86 & 3.4 & 263 & 344 & 1.0 &  157 & 170 & 0.75     \\
6kg pit &  9.671/10   & 6.202 & 6.86 & 6 & 254 & 338 & 0.99 & 159 & 172 &  0.75 \\
 \hline
\end{tabular}
\caption{The various pit and encrypting filter geometries simulated.  Statistical comparisons are performed relative to the scenario in the first row. The statistical $\chi^2$ tests were performed for the 0-60 eV and 0-200eV energy ranges, with the probabilities ($p$ values) indicating a statistically identical signals.}\label{tab:materials}
\end{table}

\begin{figure}
{\bf a} \\
    \includegraphics[width=1\textwidth]{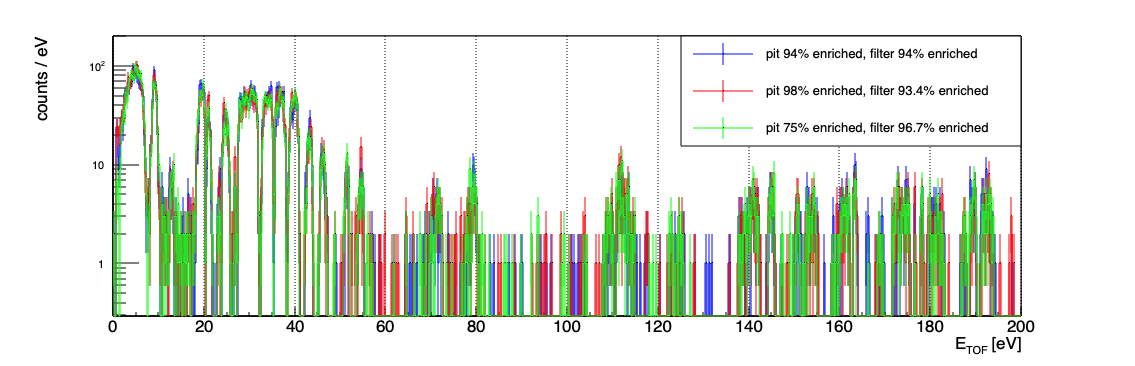}
{\bf b} \\
    \includegraphics[width=1\textwidth]{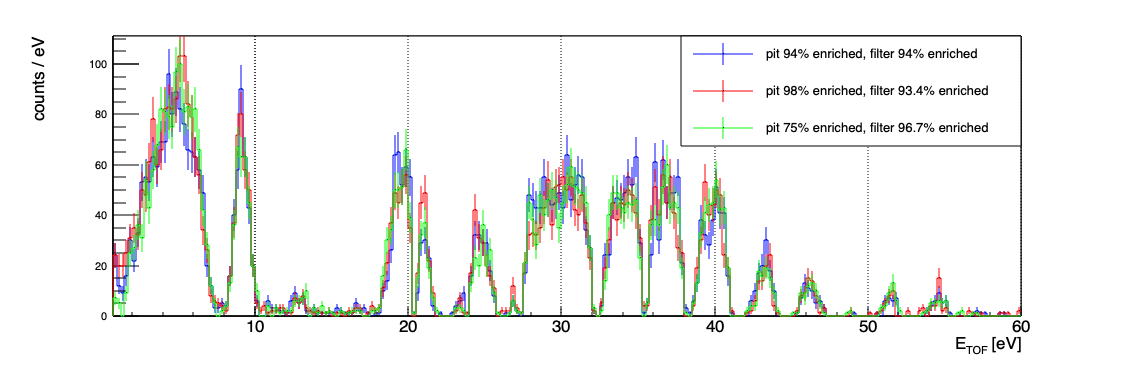}
    \caption{Results of the simulation for various isotopic enrichments.  The spectra for the three scenarios are overlaid for 0-200 eV ({\bf a}) and 0-60 eV ({\bf b}) ranges, showing overall statistically identical distributions.  The $\chi^2$ test results can be found in Supplementary Table~\ref{tab:isotopic}.}
    \label{fig:isotopic}
\end{figure}

The principle of mass encryption also applies to the isotopics of the pit.  While the inspectors may be able to infer the combined, average isotopic enrichment of the pit and the encrypting filter, that average translates to a very large range of possible pit-specific enrichment values.  To prove this, we performed simulations for a variety of enrichment levels for the pit and the filter.  The results show that given a freedom of variation on the filter enrichment, the pit enrichment may vary in the range of 75-98\%, yielding statistically identical result, as presented in Supplementary Table~\ref{tab:isotopic} and Supplementary Figure~\ref{fig:isotopic}. If the signal is independent of the pit enrichment then the signal cannot be used to infer it.  Hence as long as the inspector doesn't know the enrichment of the filter they cannot make inferences about the isotopic enrichment of the pit.   At the same time the technique shows that changing the pit enrichment from  93\% to the 60\%, typical for reactor grade plutonium (RGPu), while keeping the encrypting filter the same, as would be the case in the actual verification exercise, would result in a statistically significant difference.  Such a large observed difference would result in the rejection of the hoax.   All simulation results reflect a 5 minute-long measurement at the experimental facility where the data of this study was acquired.   At the same time, the simulations of hoaxing scenarios -- one involving a shift in isotopics, and another involving a shift in geometry -- are readily detectable, as can be seen in Supplementary Figure~\ref{fig:hoaxes} and tabulated in Supplementary Table~\ref{tab:isotopic}.   This result shows that a 5~min long measurement is easily sensitive in isotopic changes of the WGPu/RGPu scale, and geometric changes of $\sim$2~mm.  A sensitivity to smaller changes can be achieved by  increasing the measurement times, the neutron beam flux (e.g. by increasing the beam current or by shortening the distance to the photoneutron source), by choosing a more efficient photoneutron source, or by selecting a larger detector.  We can use the results from the above simulations to estimate the experimental conditions necessary for achieving diversions involving mass removals as small as 0.1~mm.  The $\chi^2$ is defined as $\chi^2=\sum (\Delta c_i/\sigma_i)^2$, where $\Delta c_i$ is the count difference in a particular bin in Supplementary Figure~\ref{fig:hoaxes} and $\sigma_i$ is the uncertainty in that difference.  Assuming that the difference changes linearly with the difference in thicknesses (true for small changes relative to the total thickness, i.e. 2$\rightarrow$0.1~mm relative to 66.8~mm),  a 20$\times$ decrease in the difference of thicknesses will translate to 20$\times$ decrease in $\Delta c_i$, in average.  This decrease in $\Delta c_i$ needs to be compensated by a similar decrease in $\sigma_i$ in order to maintain the same statistical significance. This then translates to a statistical volume which needs to be 400$\times$ larger.  For the particular experimental facility used in this study, this would mean the following changes:  bring the measurement point from the 15~m flight path distance to 5 meters, which will increase the neutron flux by 9$\times$; increase the electron beam current by 2$\times$; increase the measurement times from 5~min to approximately 2~hrs.

\begin{figure}
    \centering
    \includegraphics[width=0.9\textwidth]{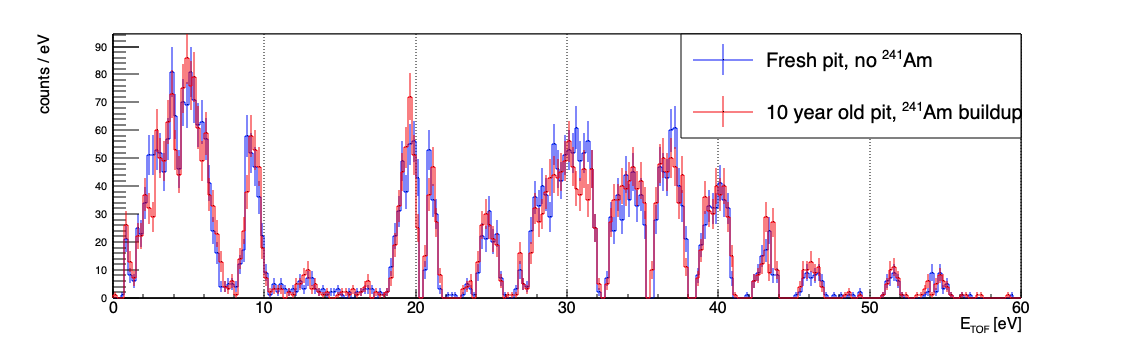}
    \caption{Comparisons of a fresh WGPu pit with no $^{241}$Am to a ten year old pit. In this simulation the isotopic ratios were taken fro Table 1 of Mark et al.\cite{ref:explosive_properties}. The overall comparison  yields a $\chi^2/NDF = 161/170$, with $p=0.68$, indicating no false positive due to the differences in americium content.}
    \label{fig:Am}
\end{figure}

A possible source of false positives may be introduced by the $\sim$14.3~year half life $\beta$-decay of $^{241}$Pu into $^{241}$Am, causing differences between honest pits of different age.  Since $^{241}$Pu constitutes only $\sim$0.3\% of WGPu \cite{ref:explosive_properties},this is a small effect, in particular given that only 8.6~mm of the combined thickness of 68.6~mm is the actual pit's thickness. For this scenario,  assuming a "fresh" filter and a pit of WGPu which is 10 years old, it can be shown that americium makes up only 1 part per 6650 of the areal density. To verify that the scale of this effect is insignificant we performed additional simulations, where we compared a "fresh" pit with no americium, to the one that has been "aged" for 10 years and has had some buildup of americium.  Supplementary Figure~\ref{fig:Am} plots the two transmission spectra.  These spectra are slightly different from the results of other simulations.  This is only due to a more precise isotopic modeling, based on Table 1 of Mark et al.\cite{ref:explosive_properties}.  The spectral comparison shows an overall agreement, and indicates that the differences of americium content should not be a cause of false positives for pits with age differences of 10 years or less.  Furthermore, the statistically identical spectra indicate that there is no americium-dependent information content in the spectra themselves.

The analysis above is relatively simple, and while some sophisticated inference analyses are theoretically possible, any concerns about information security - as well as the necessary fixes - related to americium and other impurities should be the focus of future research, and are outside of the scope of this study.  For example, the use of a chopper, as described in the main body could be used to physically limit the spectrum to very specific ranges which do not include the americium lines.  Also, the hosts can use filters with high content of americium, thus statistically “diluting” the americium-specific differences between genuine pits.   Finally, the hosts may simply choose to use this technique to dismantle weapons of similar age as the golden copy.  All these considerations and trade studies are part of the future research, and should be done with the classified knowledge of the weapons age, structure, and composition.

\begin{figure}
{\bf a} \\
    \includegraphics[width=1\textwidth]{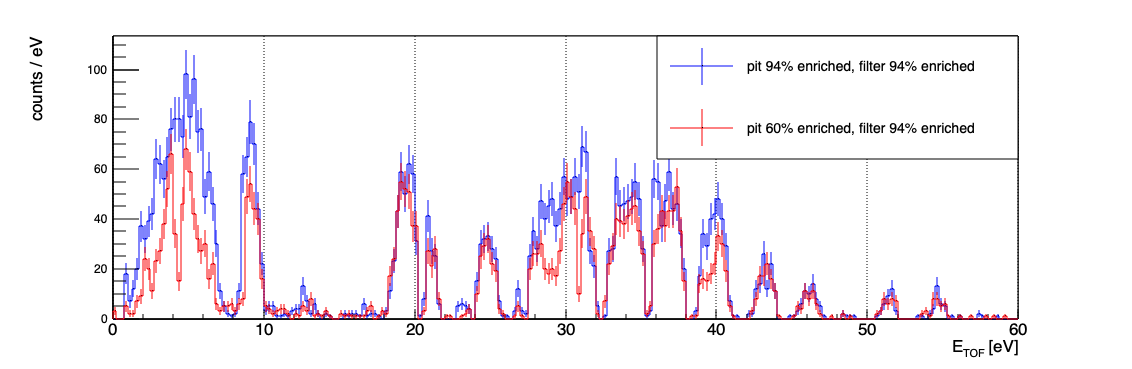}
{\bf b} \\
    \includegraphics[width=1\textwidth]{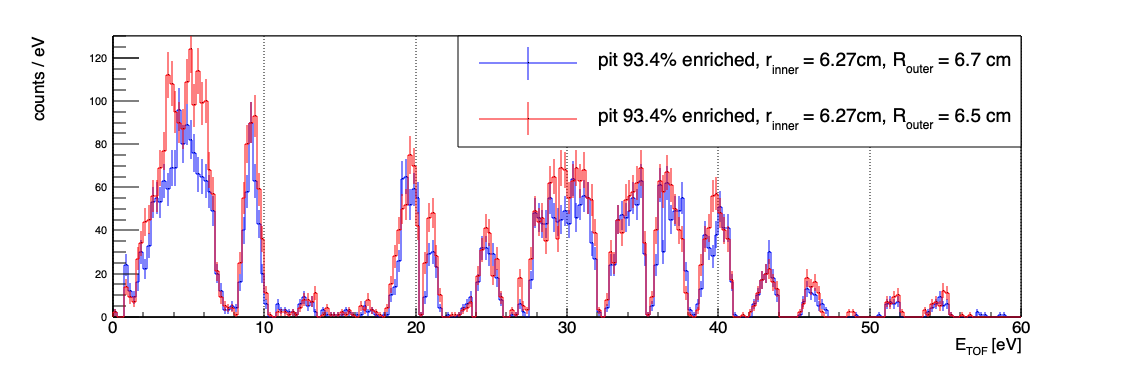}
    \caption{ Results of the simulation for two hoaxing scenarios -- isotopic replacement, and geometric change.  A replacement of the WGPu pit with a RGPu while holding the filter at WGPu enrichment readily reveals the hoax ({\bf a}).  Technique can also detect hoaxes involving correct isotopics but of a 2~mm smaller size ({\bf b}). The $\chi^2$ test results can be found in Supplementary Table~\ref{tab:isotopic}. }
    \label{fig:hoaxes}
\end{figure}

\begin{table}
\begin{tabular}{llllllllll}
\hline
            & \multicolumn{3}{c}{enrichment}               &  \multicolumn{3}{l}{0-200 eV} & \multicolumn{3}{l}{0-60 eV} \\
description & pit  & filter  & combined & $\chi^2$ & NDF &$p$ &   $\chi^2$ & NDF & $p$  \\ \hline
WGPu & 94\% & 94\% & 94\% & - & - & - & - & - & - \\
super grade & 98\% & 93.4\% & 94\% & 274 & 342 & 0.99 & 173 & 169  & 0.40 \\
low grade & 75\% & 96.7\% & 94\% & 253 & 330 & 1.0 & 158 & 165 & 0.64 \\
{\bf RGPu hoax} & 60\% & 94\%  & {\bf 89.7\%} & 698 & 342 &$\mathbf{\sim10^{-26}}$ & 573 & 169 & ${\bf \sim10^{-45}}$ \\ 
{\bf Size hoax} & 94\% & 94\%  & 94\% & 399 & 340 & {\bf 0.015} & 291 & 171 & $\mathbf{ \sim10^{-8}}$ \\ 
\hline

\end{tabular}
\caption{The various pit and encrypting filter enrichments simulated.  Statistical comparisons are performed to the scenario in the first row.  The results of the simulation indicate that as long as the combined enrichment is constant, there is no statistically significant difference in the signal.  At the same time a RGPu hoax is readily exposed. The size hoax, which involves a pit of the correct enrichment but of an outer radius 2~mm smaller than the reference object, is also detected.}\label{tab:isotopic}
\end{table}

\section*{Supplementary Note 4}

The method described in this study uses nuclear resonance phenomena as a basis for physical cryptography.  The main strength of resonance phenomena is that it involves signals that are unique to individual isotopes, allowing to achieve a one to one map from the measurement space to isotope space.  However in order for this map to be truly one to one, the resonance energies for  uranium and plutonium need to be unique to their isotopes.  While this study uses the broad energy range of [0,200] eV as part of the proof of concept demonstration, the plutonium and uranium isotopes' strongest and most clearly separated transitions are located in the narrower [0,10] eV range.  This is also the range where TOF techniques produce the highest resolution.  

We thus focus on  [0,10] eV range for a test of uniqueness.  This means searching for possible duplication of uranium and plutonium resonances in other, commonly available elements.  Supplementary Table~\ref{tab:table_of_resonances} lists the major isotopes of the elements which have resonances in the above mentioned range.  The data was acquired from the National Nuclear Data Center(NNDC)\cite{ref:nndc}, and is sorted by resonance energy.  

\begin{table}
    \centering
    \includegraphics[width=0.9\textwidth]{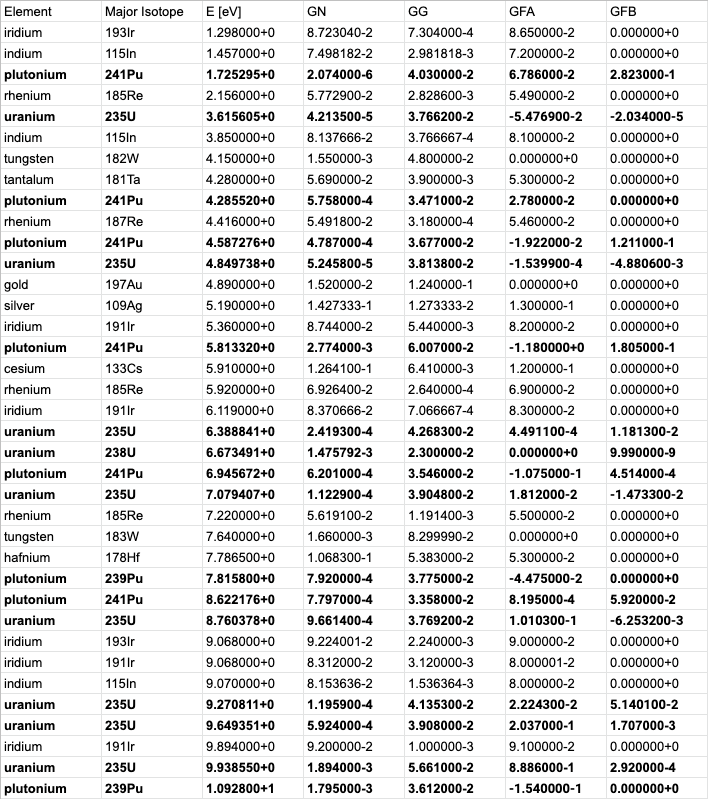}
    \caption{A listing of elements and isotopes with resonances sorted by energy. The energy range is $1$~eV~$\leq E \leq 11$~eV.  The relevant isotopes are boldfaced for comparison.  Data has been acquired from and is in the format of NNDC\cite{ref:nndc}.  GN and GG refer to the neutron scattering and capture widths, in eV.  GFA and GFB refer to the fission widths, in eV, as defined by Reich-Moore multilevel resonance formalism. }
    \label{tab:table_of_resonances}
\end{table}

An inspection of the table reveals that there are many isotopes that have resonances in this range.  Furthermore, some of these isotopes have resonances that are close in energy to some plutonium or uranium resonances. However in order to achieve an isotopic hoax, it is not sufficient to replicate an individual line.  The elements or the combination of elements in a potential hoaxing mix need to have a combination of lines that simultaneously reproduce  plutonium's resonances' energies, widths, and relative strengths.  At the same time this combination of elements should not give rise to any additional lines.  For example, to emulate $^{241}$Pu and $^{239}$Pu lines at 1.73, 4.3, 4.6, 5.8, 6.9, 7.8, 8.6, and 10.1 eV the host could try to use the 1.46, 4.3, 4.4, 5.91, 7.2, 7.8, 9.06, and 9.9 eV lines of indium, tantalum, rhenium, cesium, and hafnium, respectively.  However, this mix of elements will also produce lines at 1.3, 3.8, 5.36, and 6.1 eV - lines that are missing in plutonium isotopes. Furthermore, at higher energies these elements manifest additional lines not seen in plutonium or uranium.  
These non-fissile lines in the hoax will induce significant differences in measurements between the hoax described above and a plutonium pit, resulting in a rejection of the hoax.

While this analysis indicates the difficulty of achieving isotopic hoaxing by combination of other elements, future research needs to focus on the problem of uniqueness, i.e. systematically proving that no combination of materials can simultaneously reproduce the combined cross section shape of plutonium's isotopes.  Since future embodiment of this technique may involve more compact experimental setups with an energy resolution that is lower than the one observed in this study, the analysis of uniqueness also needs to take into account energy resolution for a particular experimental setting.

Note that Supplementary Table~\ref{tab:table_of_resonances} only includes elements that are easily available and are not difficult to handle.  The table does not list the resonances of $^{241}$Am, for example, as americium is too radioactive to be used in any hoaxing manipulation that requires $\sim$kg quantities of materials.

\section*{Supplementary References}
\bibliography{bibliography}

\bibliographystyle{naturemag} 